\documentclass[a4paper,11pt]{article}

\pdfoutput=1

\usepackage{jcappub}

\usepackage{color} 
\definecolor{Blue}{rgb}{0,0,0.9}
\definecolor{Red}{rgb}{0,0,0.9}
\definecolor{Green}{rgb}{0,0,0.9}
\definecolor{orange}{rgb}{0.8, 0.33, 0.0}
\usepackage{soul}

\usepackage[T1]{fontenc}
\usepackage{widetext}
\usepackage{subfig}
\usepackage{booktabs}
\subheader{DESY 20-040}
\title{\boldmath The Halo Void (Dust) Model of Large Scale Structure}

\author[a,b]{Rodrigo Voivodic,}
\author[c]{Henrique Rubira,}
\author[a]{Marcos Lima}

\affiliation[a]{ Departamento de F\'{\i}sica Matem\'atica, Instituto de F\'{\i}sica, Universidade de S\~ao Paulo,\\ R.  do  Mat\~ao  1371,  05508-090,  S\~ao Paulo, SP, Brazil}
\affiliation[b]{ Max-Planck-Institut f\"ur Astrophysik, Karl-Schwarzschild-Str. 1, 85741 Garching, Germany}
\affiliation[c]{Deutsches Elektronen-Synchrotron DESY, Notkestra\ss e 85, D-22607 Hamburg, Germany}

\emailAdd{rodrigo.voivodic@usp.br}
\emailAdd{henrique.rubira@desy.de}
\emailAdd{mlima@if.usp.br}

\abstract{Within the Halo Model of large scale structure, all matter is contained in dark matter halos. This simple yet powerful framework has been broadly applied to multiple data sets and enriched our comprehension of how matter is distributed in the Universe. In this work we extend this assumption by allowing for matter to rest not only inside halos but also within cosmic voids and in between halos and voids (which we call 'dust'). This assumption leads to additional contributions (1Void, 2Void, Halo-Void, etc.) to the predictions of correlation functions, spectra and profiles for both halos and voids.
Whereas the Halo Model can only make predictions for halo quantities, the Halo Void Model extends those for void statistics and halo-void cross-correlations. 
We provide recipes for all new ingredients of the Halo Void (Dust) Model, such as the void abundance, linear bias and density profile and test their validity in a N-body simulation.
Including voids and dust into the calculations improves the transition between the 1Halo and the 2Halo terms by up to $\sim 6\%$. It also eliminates the need to include low-mass structures on the normalization of large-scale terms, suggesting that halos and voids are complementary cosmic structures to effectively describe matter distribution on large scales of the Universe. 
}

\begin{document}
\maketitle
\flushbottom

\section{Introduction}
\label{sec:introduction}

Next decade large scale structure (LSS) surveys will bring the opportunity to probe cosmology in different regimes \cite{Abbetal,Amendola:2012ys,Ivezic:2008fe}. The ability to constrain models from data requires understanding in detail the origin of initial matter fluctuations in the Universe and how they evolved to form larger structures. On top of this, we must connect the underlying dark matter field to direct observables (galaxies, baryons, lensing effects, etc.). All these aspects represent difficult challenges for model Cosmology \cite{vanDaalen:2011xb,Springel2}. 

On very large scales ($\sim 100$ Mpc h$^{-1}$), matter is well described by the collisionless fluid approximation \cite{Bernardeau}. However, at these scales the small amounts of data available make it hard to overcome sample variance and constrain cosmology with high precision \cite{Abramo2, Abramo3}. On small scales, on the other hand, gravity-induced non-linear effects are 
difficult to predict from a theory standpoint, and baryonic physics effects become relevant \cite{Chisari:2019tus}. Dark matter numerical simulations \cite{Teyssier, Springel, Schneider2} can address the problem of solving non-linear scales physics. Recent codes can also evolve the baryonic fluid as well, taking into account multiple astrophysical processes  \cite{Springel2}. Despite significant improvements in simulations over the last years, they are still computationally expensive and do not provide an {\it analytical} understanding of the dependence of direct observables on cosmological parameters and initial conditions. In addition, cosmic emulators \cite{Takahashi, Heitmann, Kwan} and simplified simulations \cite{Tassev, Monaco, Bond2, Voivodic2} have been developed in order to avoid running very time-consuming N-body simulations.

Meanwhile, analytical schemes have improved our understanding of intermediate scales. The so-called Standard Perturbation Theory (SPT) constructs an expansion series over the truncated system of Navier-Stokes equations \cite{Bernardeau}, but it does not incorporate ultra-violet (UV) contributions. The Effective Field Theory (EFT) approach fixes the SPT problems incorporating UV corrections through counter-terms to be fitted to simulations \cite{Baumann:2010tm,Carrasco,Carrasco2}. Despite much progress, recent calculations of EFT expansion at three loops \cite{Konstandin:2019bay} strongly suggest that it behaves as an asymptotic series, setting a threshold to the reach of the theory at  $k \sim 0.45$ Mpc$^{-1} h$ at $z=0$.

Few analytical proposals to describe the dark matter non-linear regime have been so successful as the Halo Model (HM) of LSS \cite{Cooray}. The HM is based on the central hypothesis that all matter particles lie within halos\footnote{This assumption is of course true if we allow halos to be as small as the smallest matter particle, but it is not strictly correct when we consider structures of larger scales as in N-body simulations.}. The immediate consequence of this assumption is that all matter N-point functions split into a sum of terms that include correlations of particles that are inside the same halo or on different halos. In the case of the 2-point function, the focus of this work, there are two contributions: one coming from two particles in the same halo (1Halo) and another coming from two particles in distinct halos (2Halo). In order to compute the 1Halo and 2Halo terms, we need three ingredients: the halo density profile, the halo mass function, and the halo-halo correlation -- which at tree level can be written in terms of the halo linear bias. The success of the HM is attributed first due to its ability to make predictions for several observables without free-parameters\footnote{There might be free coefficients in the HM ingredients, e.g. the mass function \cite{Tinker}, but none of them are fitted explicitly in observables predicted by the HM.}; second because it gives an explicit separation of scales where either the 1Halo or the 2Halo terms are relevant; third because it introduces an important pedagogical framework, putting halos as central building blocks to describe matter distribution in the Universe. 
The HM also allows us to compute other quantities, such as covariance matrices \cite{Cooray, Cooray2, Lacasa}, weak-lensing correlations \cite{Kainulainen, Giocoli}, the observed halo density profile \cite{Beraldo, Lopes2}, $21$ cm and Ly{\ensuremath{\alpha}} correlations \cite{Feng3}. Moreover, the HM is useful to make predictions in models beyond $\Lambda$CDM \cite{Li11, Schmidt09, Schmidt10, Thomas}, and is the basis for the Halo Occupation Distribution (HOD) method, used e.g. to construct realistic galaxy catalogues from halo catalogues \cite{Berlind, Berlind2, Zheng}, and serving also as a basis for semi-analytical methods \cite{Takahashi}.   

Despite its triumphs, it is important to point out some of the HM deficiencies. In the transition between the 1Halo and the 2Halo terms, the agreement with simulation is only at the $\sim 20\%$ level, which at present is relatively imprecise when compared e.g. to EFTs predictions. Another problem associated with the HM is that it either requires a normalization in the 2Halo term, or the halo abundance needs to be integrated down to very low halo masses (e.g. $M \ll  10^5 M_\odot/h$ for the Press-Schechther mass function and linear bias) in order to account for all matter in the Universe. Several works have tried to modify the HM \cite{Mead, Mohammed, Seljak, Schmidt2, Chen, Valageas4}. However they either introduce new free parameters that cannot be fitted using only halo properties, or do not significantly improve the HM predictions. 

\begin{figure}
\centering
    \includegraphics[width=0.49\textwidth]{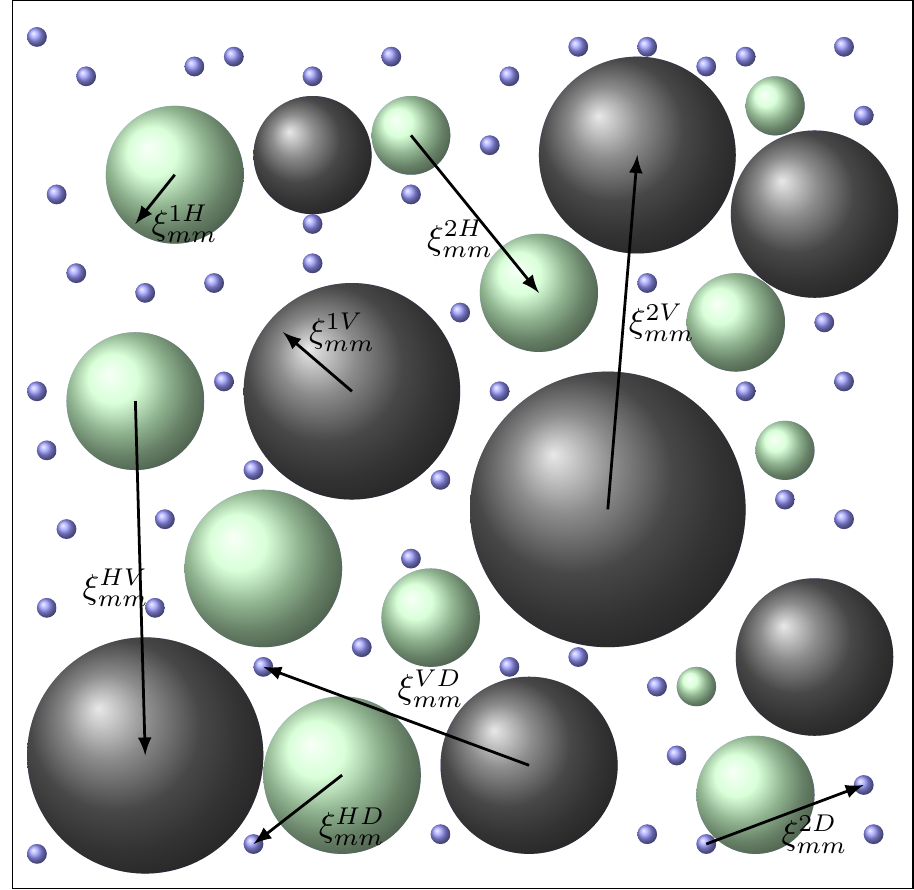} 
\caption{\label{fig:scheme}%
\small Schematic illustration of all terms contributing to the matter-matter correlation $\xi_{mm}$ within Halo Void (Dust) Model. Green spheres represent halos, gray spheres represent voids and the small blue dots represent dust particles. All terms contributing to the matter-matter correlation are indicated by arrows. Within the standard Halo Model, we have only $\xi^{1H}_{mm}$ and $\xi^{2H}_{mm}$. Considering the Halo Void Model, we add terms $\xi^{1V}_{mm}$, $\xi^{2V}_{mm}$ and $\xi^{ HV}_{mm}$. Finally, for the most general Halo Void Dust Model, we add $\xi^{HD}_{mm}$, $\xi^{VD}_{mm}$ and $\xi^{2D}_{mm}$.
}
\end{figure}

In this work, we propose two modifications in the HM central assumption mentioned previously.  
These give rise to two generalizations of the HM, namely (see Fig.~\ref{fig:scheme}):
\begin{enumerate}
    \item The Halo Void Model (HVM): matter can be distributed either in halos or in voids.
    \item The Halo Void Dust Model (HVDM): matter can be distributed in halos, voids, and in the space between them. We assume that the latter is well described by the linear regime and refer to it as 'dust'. 
\end{enumerate}

In the first scenario, the matter correlations/spectra gain contributions from five different terms: the 1Halo and 2Halo terms, already present in the HM, and the new terms 1Void, 2Void and Halo-Void. For the second scenario, dust has no internal structure (profile) so that three additional terms are needed (Void-Dust, Halo-Dust and Dust-Dust).
In principle, one could also consider contributions from other structures (e.g. filaments or sheets). We choose to stick only with voids because they can be modeled using spherical symmetry with reasonably good precision. Likewise, for voids there is a repertory of fitting prescriptions for their mass function, bias and profile. Moreover, it is easy to include voids through the excursion set formalism, whereby voids correspond to $\lesssim 40 \%$ of the total matter fraction. %

The idea of including voids as a complementary structure in the HM is in tandem with other works that enlarged the perspectives of extracting relevant information with voids. Void properties are potentially helpful in constraining cosmology
\cite{Cai2, Chantavat, Chuang, Hamaus3, Hamaus4, Hamaus5, Lavaux, Liang, Mao, Melchior, Sanchez}, primordial non-gaussianities \cite{DAmico, Chan3, Song}, neutrino properties \cite{Massara, Zhang, Sahlen2, Kreisch, Massara:2020pli}, dark energy \cite{Biswas, Pollina, Sahlen, Sutter2, Nadathur4} and modified gravity \cite{Barreira3, Cai3, Perico, Pisani, Voivodic1, Zivick, Zivick2}.  

For the HV(D)M we need three additional ingredients: the void abundance, the void density profile and the void linear bias. 
In this work we provide prescriptions and fits for each one of these ingredients using a theoretical framework and contrasting these to N-body simulation results. In the case of halo and void bias, we provide a new linear bias prescription. We also give a new prescription for the void profile with a single free parameter.
We achieve this by constructing the abundance and bias for both halos and voids simultaneously using two barriers in the excursion set formalism. This is important because the separation of particles inside halos and voids is something that does not appear as a free parameter, but as a consequence of the excursion set theory with a double barrier. The HV(D)M is fully self-consistent, as it also takes into account the void-in-cloud, void-in-void and cloud-in-void effects, analogously to the so-called cloud-in-cloud effect that appears when considering only halos.
 In addition, we show that using two barriers eliminates the need for a normalization on large scales or the need to extrapolate the abundance down to tiny halo masses (in contrast to the usual HM prescription). This happens in our generalized models because the matter within small halos is taken into account within larger voids.  This highlights the fact that halos and voids are complementary structures which, when considered together, give a better effective description of how matter is distributed in the Universe. 

In this work we present results for multiple observables: the matter-matter auto power spetrum, the halo-matter, void-matter cross-spectra as well as the halo and void density profiles. For the matter-matter power spectrum, we show that the HVM (HVDM) improves the transition from the 2Halo to the 1Halo term by $\approx 2\%$ ($\approx 6 \%$), when compared to the standard HM. This happens because the typical void radius scale is larger than the typical halo radius -- even though their masses are smaller -- which adds a new scale acting in the transition from the 1Halo to the 2Halo term, alleviating the lack of power in these scales. Even though this improvement is somewhat marginal, we show that the HVM a correct approach to model the matter--matter, halo--matter and void--matter observables in a self-consistent way.

This work is structured in the following way. In Sec.~\ref{sec:halo_void_model} we present the main set up for both the HVM and the HVDM. In Sec.~\ref{sec:halovoid_ingredients}, we provide prescriptions for each of the ingredients needed --  profiles, linear bias and mass functions. We also how the complementarity of voids and halos solves the large-scale normalization problem of the HM. We give new prescriptions for the void profiles and for the halo and void linear bias considering two linear diffusive barriers within the excursion set formalism. 
In Sec.~\ref{sec:results}, we present results for the matter-matter power spectrum, as well as the cross-spectra and the halo and void profiles. In Sec.~\ref{sec:conclusion}, we conclude and in App.~\ref{app:abundance_bias} we present computations of the mass function and linear bias using the excursion set theory. 

\section{The Halo Void (Dust) Model}
\label{sec:halo_void_model}

After reviewing the standard HM of large scale structure \cite{Cooray}, 
we construct the HVM and indicate the additional ingredients needed. We then describe the HVDM, including one more component: matter particles that are neither in halos nor voids, which we label 'dust'. In all scenarios we provide explicit corrections for matter-matter auto-correlations as well as cross-correlations (halo-matter and void-matter).

\subsection{The Halo Model}\label{sec:hm}

The fundamental HM assumption is that all matter in the Universe belongs to dark matter halos \cite{Cooray}. This implies that the matter density field can be written as the sum of the contributions of matter within the density profiles of all halos in the Universe
\begin{equation}
\rho (\textbf{x}) = \sum _{i}^{\rm halos} \rho _{h}(\textbf{x} - \textbf{x}_{i}|M_{i}) \,,
\label{eq:Density_field}
\end{equation}
where $\rho _{h}(\textbf{x} - \textbf{x}_{i}|M_{i})$ is the matter density profile of a halo centered at position $\textbf{x}_{i}$ with  mass $M_{i}$.
We may rewrite Eq.~\eqref{eq:Density_field} as
\begin{equation}
\rho (\textbf{x}) = \sum_{i}^{\rm halos} \int d M \, \delta _{D}(M - M_{i}) \int d^{3}x' \, \delta _{D}(\textbf{x}' - \textbf{x}_{i}) M u_h(\textbf{x} - \textbf{x}'|M) \,,
\label{eq:Density_rewrite}    
\end{equation}
where $\delta _{D}(\textbf{x})$ is the 3D Dirac's delta distribution and $u_h(\textbf{x}|M) = \rho _{h}(\textbf{x}|M)/M$ is the mass-normalized halo density profile.

From Eq.~\eqref{eq:Density_rewrite} we can write predictions for the matter correlation function and its Fourier counterpart, the matter power spectrum. For the two-point correlation function of the matter density field we have
\begin{eqnarray}
\left \langle \rho (\textbf{x}) \rho (\textbf{x} + \textbf{r})\right \rangle &=& \int dM_{1} \int dM_{2} \int d^{3} x_{1} \int d^{3} x_{2} \, M_{1} M_{2} u_h(\textbf{x} - \textbf{x}_{1}|M_{1}) u_h(\textbf{x} - \textbf{x}_{2} + \textbf{r}|M_{2}) \nonumber \\
&\times & \left \langle \sum _{i, j} \delta _{D}(M_{1} - M_{i}) \delta _{D}(M_{2} - M_{j})\delta _{D}(\textbf{x}_{1} - \textbf{x}_{i}) \delta _{D}(\textbf{x}_{2} - \textbf{x}_{j}) \right \rangle \,.
\label{eq:Density_two}
\end{eqnarray}
The expectation value in Eq.~\eqref{eq:Density_two} can be split into two terms, $I^{1H}$ and $I^{2H}$. The first contains the contribution of terms in the sum with $i = j$, such that the matter particles are within the same halo (1Halo). The second contribution comes from terms with $i\neq j$ and refers to matter particles from different halos (2Halo). These two terms can be written as
\begin{eqnarray}
I^{1H} &=& \delta _{D}(M_{1} - M_{2})\delta _{D}(\textbf{x}_{1} - \textbf{x}_{2}) \frac{dn_h}{dM_{1}} \,,\label{eq:Is1} \\
I^{2H} &=&  \frac{dn_h}{dM_{1}} \frac{dn_h}{dM_{2}} \left[ 1 + \xi _{hh} (\textbf{x}_{1} - \textbf{x}_{2}| M_{1}, M_{2}) \right] \,,
\label{eq:Is}
\end{eqnarray}
where $dn_h/dM$, the halo mass function,  measures the differential number density of halos in mass bin $[M, M + dM]$ and $\xi _{hh} (\textbf{x}_{1} - \textbf{x}_{2}| M_{1}, M_{2})$ is the halo-halo two-point correlation of halos with masses $M_{1}$ and $M_{2}$.
In order to obtain Eq.~\eqref{eq:Is} we used
\begin{equation}
\left \langle \sum _{i} \delta _{D}(M_{1} - M_{i}) \delta _{D}(\textbf{x}_{1} - \textbf{x}_{i}) \right \rangle = \frac{dn_h}{d M_{1}} \,,
\end{equation}
which
follows from setting the background matter density $\bar {\rho}_{m} \equiv \left \langle \rho (\textbf{x}) \right \rangle $ and the fact that all matter in the Universe is within halos so that
\begin{equation} \label{eq:hm_constraint}
\bar{\rho}_{m} = \int dM \, M \frac{dn_h}{dM} \,.
\end{equation}

\subsubsection*{Matter-matter auto-correlation}

Using Eqs.~\eqref{eq:Is1} and \eqref{eq:Is} the matter-matter two-point correlation function can be written as
\begin{equation}
\xi_{mm}(\textbf{r}) = \frac{1}{\bar{\rho} ^{2}_{m}} \left[  \left \langle \rho (\textbf{x}) \rho (\textbf{x} + \textbf{r})\right \rangle - \bar{\rho} ^{2}_{m}\right] = \xi_{mm}^{1H} (\textbf{r}) +  \xi_{mm}^{2H}(\textbf{r}) \,,
\label{eq:Ximm}
\end{equation}
where
\begin{eqnarray} \label{eq:a_1h_mm}
\xi_{mm} ^{1H} (\textbf{r}) &=& \int dM_{1} \int dM_{2} \int d^{3} x_{1} \int d^{3} x_{2} \delta _{D}(M_{1} - M_{2})\delta _{D}(\textbf{x}_{1} - \textbf{x}_{2}) \nonumber \\
&\times & \frac{dn_h}{dM_{1}} \frac{M_{1}}{\bar{\rho} _{m}} \frac{M_{2}}{\bar{\rho} _{m}} u_h(\textbf{x} - \textbf{x}_{1}|M_{1}) u_h(\textbf{x} - \textbf{x}_{2} + \textbf{r}|M_{2}) \nonumber \\
&=& \int d\ln M  \frac{M^{2}}{\bar{\rho}^{2} _{m}}\frac{dn_h}{d\ln M} \int d^{3} y \, u_h(\textbf{y}|M) u_h( \textbf{y} + \textbf{r}|M) \,,
\end{eqnarray}
and
\begin{eqnarray}\label{eq:a_2h_mm}
\xi_{mm}^{2H} (\textbf{r}) &=& \int d\ln M_{1} \frac{M_{1}}{\bar{\rho} _{m}} \frac{dn_h}{d\ln M_{1}} \int d\ln M_{2} \frac{M_{2}}{\bar{\rho} _{m}} \frac{dn_h}{d\ln M_{2}} \nonumber \\
&\times& \int d^{3} y_{1} \, u_h(\textbf{y}_{1}|M_{1}) \int d^{3} y_{2} \, u_h(\textbf{y}_{2} + \textbf{r}|M_{2}) \xi _{hh} (\textbf{y}_{1} - \textbf{y}_{2}| M_{1}, M_{2}) \,.
\end{eqnarray}

Assuming spherical symmetry in the halo density profiles, i.e. $u_h(\textbf{r}|M) =u_h(r|M)$, and Fourier transforming Eq.~\eqref{eq:Ximm}, the matter-matter power spectrum becomes
\begin{equation}
P_{mm}(k) = P_{mm}^{1H}(k) + P_{mm}^{2H}(k)\,,
\label{eq:Halo_Model}
\end{equation}
where 
\begin{equation}\label{eq:P1halo}
P_{mm}^{1H}(k) = \int d \ln M \, \frac{M^{2}}{\bar{\rho}^{2}_{m}} \frac{dn_h}{d \ln M}|u_h(k|M)|^{2} \,,
\end{equation}
and
\begin{equation}\label{eq:P2halo}
P_{mm}^{2H}(k) = \int d \ln M_{1} \, \frac{M_{1}}{\bar{\rho}_{m}} \frac{dn_h}{d \ln M_{1}}u_h(k|M_{1}) \int d \ln M_{2} \, \frac{M_{2}}{\bar{\rho}_{m}} \frac{dn_h}{d \ln M_{2}}u_h(k|M_{2}) P_{hh}(k|M_{1}, M_{2}) \,.
\end{equation}
Here $u_h(k|M)$ is the Fourier transformation of the normalized density profile and $P_{hh}$ is the halo-halo power spectrum. 

At this point, it is common to consider the linear approximation for $P_{hh}(k|M_1, M_2)=b_h^L(M_1)b_h^L(M_2)P_{mm}^L(k)$ \cite{Cooray}. Under this approximation, the 2Halo term becomes
\begin{equation} \label{eq:2haloterm}
P_{mm}^{2H}(k) = \left[ \int d \ln M \, \frac{M}{\bar{\rho}_{m}} \frac{dn_h}{d \ln M}u_h(k|M) b^{L}_{h}(M) \right]^{2} P^{L}_{mm}(k)\,,
\end{equation}
where $b^{L}_{h}(M)$ is the linear halo bias and $P^{L}_{mm}(k)$ is the matter-matter linear power spectrum.
Note that, on large scales, $u_h(k|M) \rightarrow 1$, implying that $P^{2H}_{mm}(k) \rightarrow P^{L}_{mm}(k)$, as long as the bias and mass function are properly normalized, i.e. the matter is not biased with respect to itself, which translates into the constraint 
\begin{equation}
\int d\ln M \, \frac{M}{\bar{\rho}_{m}} \frac{dn_{h}}{d\ln M} b_{h} ^{L}(M) = 1 \,.
\end{equation}
However, depending on the halo mass function and bias used, this integral has a very slow convergence in the lower limit.  In order to properly normalize the 2Halo term, it is typically necessary to integrate down to very small halo masses, which requires extrapolating the mass function and the linear bias beyond the ranges in which they are calibrated from simulations. As we will show in Sec.~\ref{sec:consistence}, this problem is solved in the context of the HVM.

\subsubsection*{Cross-correlation}
Another prediction of the HM is the halo-matter two-point correlation function. Proceeding similarly to the matter-matter case, we can write
\begin{equation}
\xi_{hm}(\textbf{r}|M) = \frac{1}{\bar{\rho} ^{2}_{m}} \left[  \left \langle \rho _{h} (\textbf{x}|M) \rho (\textbf{x} + \textbf{r})\right \rangle - \bar{\rho} ^{2}_{m}\right] = \xi_{hm}^{1H}(\textbf{r}|M) +  \xi_{hm}^{2H} (\textbf{r}|M) \,,
\label{eq:Xihm}
\end{equation}
where we used $\xi_{hm}$ for each term contributing to the cross-correlation. 
By a similar procedure as done for $\xi_{mm}$ one can calculate each term above as 
\begin{eqnarray}
\xi_{hm}^{1H}(\textbf{r}|M) &=& \frac{\rho _{h}(\textbf{r}|M)}{ \bar{\rho}_{m}} \,, \label{eq:a_hm_1h}\\
\xi_{hm}^{2H}(\textbf{r}|M) &=& \int d \ln M_{1} \, \frac{M_{1}}{\bar{\rho}_{m}} \frac{dn_h}{d \ln M_{1}} \int d^{3} y \, u_h(\textbf{y} + \textbf{r}|M_{1}) \xi_{hh}(\textbf{y}|M_{1}, M) \label{eq:a_hm_2h}\,.
\end{eqnarray}

Assuming again spherical symmetry for the profile, in Fourier space we have for the halo-matter cross-spectrum 
\begin{equation} \label{eq:HM_phm}
    P_{hm}(k|M) = P_{hm}^{2H}(k|M) + P_{hm}^{1H}(k|M), 
\end{equation}
The 1Halo and 2Halo cross-correlations are given by
\begin{equation}
P_{hm}^{1H}(k|M) = \frac{M}{\bar{\rho}_m} u_h(k|M) \,,
\end{equation}
and 
\begin{equation}
P_{hm}^{2H}(k|M) = \int d \ln M_{1} \, \frac{M_{1}}{\bar{\rho}_{m}} \frac{dn_h}{d \ln M_{1}} u_h(k|M_{1}) P_{hh}(k|M_{1}, M) \,.
\label{eq:Halo_matter_2h}
\end{equation}
On scales, where linear theory is valid, $P_{hm}^{2H}$ becomes
\begin{equation}
P_{hm}^{2H}(k|M) = \left[\int d \ln M_{1} \, \frac{M_{1}}{\bar{\rho}_{m}} \frac{dn_h}{d \ln M_{1}} u_h(k|M_{1}) b^{L}_{h}(M_{1}) \right] b^{L}_{h}(M) P^{L}_{mm}(k) \,.
\end{equation}
For large scales where $ u(k|M_{1}) \approx 1$, we have
\begin{equation}
P_{hm}^{2H}(k|M) = b^{L}_{h}(M) P^{L}_{mm}(k) \,,
\end{equation}
and finally
\begin{equation}
\xi _{hm} (r|M) = \frac{\rho _{h}(r|M)}{ \bar{\rho}_{m}} +  b^{L}_{h}(M) \xi ^{L}_{mm} (r)\,\label{eq:xi_profile}.
\end{equation}
This expression has been used to describe the halo density profile observed beyond the 1Halo contribution \cite{Beraldo, Lopes2}. For the crossed-statistics, it is also important to include an exclusion term that suppresses the second term of Eq.~(\ref{eq:xi_profile}) inside the halo. We comment more on this exclusion term in Sec.~\ref{sec:bias}.

In summary, within the HM, we need prescriptions for the halo mass function, the halo linear bias and the halo density profile in order to predict non-linear correlation functions.
Despite the HM success, it has a few deficiencies: the matter power spectra is underestimated in the transition between the 1Halo and 2Halo terms, and the mass function and linear bias need to be extrapolated down to very small masses. 
In the next subsection, we propose a modification of the HM, which incorporates matter particles of low massive halos into cosmic voids.  

\subsection{The Halo Void Model}
\label{sec:hv_model}

We now define the HVM, by changing Eq.~\eqref{eq:Density_field}, the HM fundamental assumption that all matter is within halos. We will now suppose that the matter can also be inside cosmic voids such that
\begin{equation}
\rho (\textbf{x}) = \sum ^{\rm halos} _{i} \rho _{h} (\textbf{x} - \textbf{x}_{i}|M_{i}) + \sum ^{\rm voids} _{j} \rho _{v} (\textbf{x} - \textbf{x}_{j}|M_{j}) \,,
\label{eq:Density_field_hv}
\end{equation}
where $\rho _{h} (\textbf{x} - \textbf{x}_{i}|M_{i})$ is the density profile of a halo with mass $M_{i}$ centered at $\textbf{x}_{i}\,$ and $\rho _{v} (\textbf{x} - \textbf{x}_{j}|M_{j})$ is the density profile of a void with mass $M_{j}$ centered at $\textbf{x}_{j}$.

Similarly to what was done in Eq.~\eqref{eq:Density_rewrite}, we can rewrite Eq.~\eqref{eq:Density_field_hv} as
\begin{eqnarray}
\rho (\textbf{x}) &=& \int dM \int d^{3} x' \left[ \sum ^{\rm halos} _{i} \delta _{D} (M - M_{i}) \delta _{D}(\textbf{x}' - \textbf{x}_{i}) M u_{h}(\textbf{x} - \textbf{x}'|M) \right.  \nonumber \\
&+& \left. \sum ^{\rm voids} _{j} \delta _{D} (M - M_{j}) \delta _{D}(\textbf{x}' - \textbf{x}_{j}) M u_{v}(\textbf{x} - \textbf{x}'|M) \right] \,.
\end{eqnarray}
\subsubsection*{Matter-matter auto-correlation}
The two point correlation function of the matter density field is now given by
\begin{equation}
\xi _{mm} (\textbf{r}) = \frac{1}{\bar{\rho}^{2}_{m}}\left[ \left \langle \rho (\textbf{x}) \rho (\textbf{x} + \textbf{r})\right \rangle - \bar{\rho}^{2}_{m} \right] = \xi_{mm}^{1H} + \xi_{mm}^{2H} + \xi_{mm}^{1V} + \xi_{mm}^{2V} + 2\xi_{mm}^{HV} \,,
\end{equation}
where $\xi_{mm}^{1H}$ and $\xi_{mm}^{2H}$ are given respectively by Eqs.~(\ref{eq:a_1h_mm}) and (\ref{eq:a_2h_mm}). The other terms can be calculated following the same computation steps in Sec.~\ref{sec:hm}, giving
\begin{eqnarray}
\xi_{mm}^{1V}(\textbf{r}) &=& \int d\ln M  \frac{M^{2}}{\bar{\rho}^{2}_{m}}\frac{dn_{v}}{d\ln M} \int d^{3} y \, u_{v}(\textbf{y}|M) u_{v}( \textbf{y} + \textbf{r}|M) \,, \\
\xi_{mm}^{2V}(\textbf{r}) &=&  \int d\ln M_{1} \frac{M_{1}}{\bar{\rho} _{m}} \frac{dn_{v}}{d\ln M_{1}} \int d\ln M_{2} \frac{M_{2}}{\bar{\rho} _{m}} \frac{dn_{v}}{d\ln M_{2}} \nonumber \\
&\times& \int d^{3} y_{1} \, u_{v}(\textbf{y}_{1}|M_{1}) \int d^{3}y_{2} \, u_{v}(\textbf{y}_{2} + \textbf{r}|M_{2}) \xi _{vv} (\textbf{y}_{1} - \textbf{y}_{2}| M_{1}, M_{2}) \,, \\
\xi_{mm}^{HV}(\textbf{r})  &=&  \int d\ln M_{1} \frac{M_{1}}{\bar{\rho} _{m}} \frac{dn_{h}}{d\ln M_{1}} \int d\ln M_{2} \frac{M_{2}}{\bar{\rho} _{m}} \frac{dn_{v}}{d\ln M_{2}} \nonumber \\
&\times& \int d^{3} y_{1} \, u_{h}(\textbf{y}_{1}|M_{1}) \int d^{3}y_{2} \, u_{v}(\textbf{y}_{2} + \textbf{r}|M_{2}) \xi _{hv} (\textbf{y}_{1} - \textbf{y}_{2}| M_{1}, M_{2}) \,.
\end{eqnarray}
Here $\xi_{vv}$ and $\xi_{hv}$ designate the void-void and halo-void correlation functions, with $dn_v/d\ln M$ describing the void abundance. The new terms in the expression above take into account contributions to the matter-matter correlation function from: two points in the same void ($\xi_{mm}^{1V}$); two points in different voids ($\xi_{mm}^{2V}$); and one point in a halo and one in a void ($\xi_{mm}^{HV}$).

In Fourier space the matter-matter power spectrum becomes
\begin{equation}
P_{mm}(k) =  P_{mm}^{1H}(k) + P_{mm}^{2H}(k) + P_{mm}^{1V}(k) + P_{mm}^{2V}(k) + 2P_{mm}^{HV}(k)  \,,
\label{eq:Halo_Void_Model}
\end{equation}
where $P_{mm}^{1H}$ and $P_{mm}^{2H}$ are given by Eqs.~\eqref{eq:P1halo} and \eqref{eq:P2halo}. For the other terms (assuming again spherical symmetry in the profiles) we have
{\small \begin{eqnarray}
P_{mm}^{1V}(k) &=& \int d\ln M \frac{M^{2}}{\bar{\rho}^{2}_{m}} \frac{dn_{v}}{d\ln M} |u_{v}(k|M)|^{2} \,, \\
P_{mm}^{2V}(k) &=& \int d\ln M_{1} \frac{M_{1}}{\bar{\rho} _{m}} \frac{dn_{v}}{d\ln M_{1}} u_{v}(k|M_{1}) \int d\ln M_{2} \frac{M_{2}}{\bar{\rho} _{m}} \frac{dn_{v}}{d\ln M_{2}}u_{v}(k|M_{2}) P_{vv}(k|M_{1}, M_{2}) \,, \label{eq:P2void}\\
P_{mm}^{HV}(k) &=& \int d\ln M_{1} \frac{M_{1}}{\bar{\rho} _{m}} \frac{dn_{h}}{d\ln M_{1}} u_{h}(k|M_{1}) \int d\ln M_{2} \frac{M_{2}}{\bar{\rho} _{m}} \frac{dn_{v}}{d\ln M_{2}}u_{v}(k|M_{2}) P_{hv}(k|M_{1}, M_{2}) \,\label{eq:Phv},
\label{eq:PHV_terms}
\end{eqnarray} }
where $P_{vv}$ and $P_{hv}$ are the Fourier transforms of $\xi_{vv}$ and $\xi_{hv}$.

Note that since now the matter in the Universe is distributed between halos and voids, we must have that
\begin{equation} \label{eq:hvm_sum}
\bar{\rho} _{m} = \bar{\rho}^{h} _{m} + \bar{\rho}^{v} _{m}\,,
\end{equation}
where $\bar{\rho} _{m}$ is the total matter density in the Universe and $\bar{\rho}^{h} _{m}$ and $\bar{\rho}^{v} _{m}$ are the contributions coming from halos and voids
\begin{eqnarray}
\bar{\rho }_{m}^{h} &\equiv & \int d\ln M \, M \frac{dn_{h}}{d\ln M}   \,, \label{eq:h_constraint} \\
\bar{\rho }_{m}^{v} &\equiv & \int d\ln M \, M \frac{dn_{v}}{d\ln M} \,. \label{eq:v_constraint}
\end{eqnarray}
We must also ensure that the matter field is not biased with respect to itself so that
\begin{equation}\label{eq:hvm_constraint}
   \bar{b}_{m}^{h} + \bar{b}_{m}^{v} = 1 \,,
\end{equation}
where $\bar{b}_{m}^{h}$ and $\bar{b}_{m}^{v}$ are the mean bias of matter in halos and of matter in voids, defined as
\begin{eqnarray} 
\bar{b}_{m}^{h} &\equiv& \int d\ln M \, \frac{M}{\bar{\rho}_{m}} \frac{dn_{h}}{d\ln M} b_{h} ^{L}(M)  \label{eq:hb_constraint} \,, \\
\bar{b}_{m}^{v} &\equiv& \int d\ln M \, \frac{M}{\bar{\rho}_{m}} \frac{dn_{v}}{d\ln M} b_{v} ^{L}(M)  \label{eq:vb_constraint} \,.
\end{eqnarray}
Using these integral constraints on large scales, where both profiles go to unity, and neglecting the 1Halo and 1Void terms, we recover the linear matter-matter power spectrum in Eq.~(\ref{eq:Halo_Void_Model}). Notice that these constraints are generalizations of the corresponding constraints in the HM. As shown in Sec.~\ref{sec:consistence}, differently from the HM, these constraints are easily satisfied by the HVM through the excursion set formalism with two barriers.

\subsubsection*{Cross-correlations}
In analogy to Eq.~\eqref{eq:Halo_Void_Model}, the halo-matter and the void-matter two-point correlation functions are given by
 \begin{eqnarray} 
\xi_{hm}(\textbf{r}, M) &=& \frac{1}{\bar{\rho}_{m} \bar{\rho}_{m}^{h}} \left[  \left \langle \rho _{h} (\textbf{x}|M) \rho (\textbf{x} + \textbf{r})\right \rangle - \bar{\rho}_{m} \bar{\rho}_{m}^{h}\right]  \nonumber \\
&=& \xi_{hm}^{1H}(\textbf{r}|M) + \xi_{hm}^{2H}(\textbf{r}|M) + \xi_{hm}^{HV}(\textbf{r}|M) \,,\label{eq:xi_profile_h} \\
\xi_{vm}(\textbf{r}, M) &=& \frac{1}{\bar{\rho}_{m} \bar{\rho}_{m}^{v}} \left[  \left \langle \rho _{v} (\textbf{x}|M) \rho (\textbf{x} + \textbf{r})\right \rangle - \bar{\rho}_{m} \bar{\rho}_{m}^{v}\right] \nonumber \\
&=& \xi_{vm}^{1V}(\textbf{r}|M) + \xi_{vm}^{2V}(\textbf{r}|M) + \xi_{vm}^{HV}(\textbf{r}|M) \label{eq:xi_profile_v}\,,
\end{eqnarray} 
where $\xi_{hm}^{1H}$ and $\xi_{hm}^{2H}$ are given respectively by Eqs.~(\ref{eq:a_hm_1h}) and (\ref{eq:a_hm_2h}). The other terms are given by
\begin{eqnarray}
\xi_{hm}^{HV}(\textbf{r}|M) &=&  \int d \ln M_{1} \, \frac{M_{1}}{\bar{\rho}_{m}} \frac{dn_{v}}{d \ln M_{1}} \int d^{3} y \, u_{v}(\textbf{y} + \textbf{r}|M_{1}) \xi_{hv}(\textbf{y}|M_{1}, M)\,, \\
\xi_{vm}^{HV}(\textbf{r}|M) &=&  \int d \ln M_{1} \, \frac{M_{1}}{\bar{\rho}_{m}} \frac{dn_{h}}{d \ln M_{1}} \int d^{3} y \, u_{h}(\textbf{y} + \textbf{r}|M_{1}) \xi_{hv}(\textbf{y}|M_{1}, M)\,, \\
\xi_{vm}^{1V}(\textbf{r}|M) &=& \frac{\rho _{v}(\textbf{r}|M)}{ \bar{\rho}_{m}} \,, \\
\xi_{vm}^{2V}(\textbf{r}|M) &=& \int d \ln M_{1} \, \frac{M_{1}}{\bar{\rho}_{m}} \frac{dn}{d \ln M_{1}} \int d^{3} y \, u_v(\textbf{y} + \textbf{r}|M_{1}) \xi_{vv}(\textbf{y}|M_{1}, M) \,.
\end{eqnarray}
Again, using the linear bias and mass functions numerical constraints and assuming spherical profiles for halos and voids, we recover the predictions at linear scales: $\xi_{hm}^{2H}(r| M) = b_{h}^{L}(M) \xi_{mm}^{L}(r)$ and $\xi_{vm}^{2V}(r| M) = b_{v}^{L}(M) \xi_{mm}^{L}(r)$. Here, once more, it is vital to include an exclusion term that kills the 2H, 2V and HV terms inside halos and voids (see Sec.~\ref{sec:bias}).

Notice that $\xi_{hm}(\textbf{r}| M)$ and $\xi_{vm}(\textbf{r}| M)$ can be interpreted as the {\it observed} halo and void profiles. The first term in Eq.~(\ref{eq:xi_profile_h}) models the inner part of the profile (e.g. NFW for the halos), while the other terms are additional contributions that improve the profile prediction in the outer regions. 

In Fourier space, the halo-matter and void-matter power spectra gain additional contributions. For instance, Eq.~(\ref{eq:HM_phm}) becomes
\begin{equation} \label{eq:HVM_phm}
    P_{hm}(k|M) = P_{hm}^{2H}(k|M) + P_{hm}^{1H}(k|M) + P_{hm}^{HV}(k|M)\,, 
\end{equation}
and for voids
\begin{equation} \label{eq:HVM_pvm}
    P_{vm}(k|M) = P_{vm}^{2V}(k|M) + P_{vm}^{1V}(k|M) + P_{vm}^{HV}(k|M)\,,
\end{equation}
with
\begin{eqnarray}
P_{hm}^{HV}(k|M) &=& \int d \ln M_{1} \, \frac{M_{1}}{\bar{\rho}_{m}} \frac{dn_v}{d \ln M_{1}} u_v(k|M_{1}) P_{hv}(k|M_{1}, M) \,, \\
P_{vm}^{HV}(k|M) &=& \int d \ln M_{1} \, \frac{M_{1}}{\bar{\rho}_{m}} \frac{dn_h}{d \ln M_{1}} u_h(k|M_{1}) P_{hv}(k|M_{1}, M) \,, \\
P_{vm}^{1V}(k|M) &=& \frac{M}{\bar{\rho}_{m}} u_v(k|M)\,, \\
P_{vm}^{2V}(k|M) &=& \int d \ln M_{1} \, \frac{M_{1}}{\bar{\rho}_{m}} \frac{dn_v}{d \ln M_{1}} u_v(k|M_{1}) P_{vv}(k|M_{1}, M) \,.
\end{eqnarray}
%

\subsection{The Halo Void Dust Model} \label{sec:hvd_model}

We now consider a variation of the HVM: in addition to halos and voids, we consider that in the space between these structures there is a significant fraction of matter that follows the linear regime. We refer to this underlying linear matter field as 'dust' and modify Eq.~(\ref{eq:Density_field_hv}) into
\begin{equation}
\rho (\textbf{x}) = \sum ^{\rm halos} _{i} \rho _{h} (\textbf{x} - \textbf{x}_{i}|M_{i}) + \sum ^{\rm voids} _{j} \rho _{v} (\textbf{x} - \textbf{x}_{j}|M_{j}) + \rho _{d}(\textbf{x}) \,,
\label{eq:Density_field_hvd}
\end{equation}
where $\rho _{d}(\textbf{x})$, the dust density, is the residual matter that follows linear theory and is not inside any structure (halos or voids).

\subsubsection*{Matter-matter auto-correlation}
In this case, the matter two-point correlation function will be given by
\begin{eqnarray}
\xi _{mm} (\textbf{r}) &=& \frac{1}{\bar{\rho}^{2}_{m}}\left[ \left \langle \rho (\textbf{x}) \rho (\textbf{x} + \textbf{r})\right \rangle - \bar{\rho}^{2}_{m} \right]  \nonumber \\ &=& \xi_{mm}^{1H} + \xi_{mm}^{2H} + \xi_{mm}^{1V} + \xi_{mm}^{2V} + 2\xi_{mm}^{HV} + 2\xi_{mm}^{HD} + 2\xi_{mm}^{VD} + \xi_{mm}^{2D} \,,
\end{eqnarray}
where terms involving dust 
are given by
\begin{eqnarray}
\xi_{mm}^{HD}(\textbf{r})  &=& \int d \ln M \frac{M}{\bar{\rho}_{m}} \frac{d n_{h}}{d \ln M} \int d^{3}y \, u_{h}(\textbf{y}|M) \xi _{hd}(\textbf{y} + \textbf{r}|M) \,, \\
\xi_{mm}^{VD}(\textbf{r})  &=&  \int d \ln M \frac{M}{\bar{\rho}_{m}} \frac{d n_{v}}{d \ln M} \int d^{3}y \, u_{v}(\textbf{y}|M) \xi _{vd}(\textbf{y} + \textbf{r}|M) \,, \\
\xi_{mm}^{2D}(\textbf{r}) &=&  \xi _{dd} (\textbf{r})  \,.
\end{eqnarray}
The new terms above quantify contributions to the matter-matter correlation function coming from the following pairs of points: one in a halo and one at a dust point ($\xi_{mm}^{HD}$); one in a void and one at a dust point ($\xi_{mm}^{VD}$); and at two different dust points ($\xi_{mm}^{2D}$). Since dust is considered to be structureless, we have $\xi_{mm}^{1D}(r) = 0$. 

As mentioned above, we consider the matter in dust to be distributed in the Universe following linear perturbations. 
As a result, correlation functions including dust are given by
\begin{eqnarray}
\xi _{hd}(\textbf{x}|M) &=& \bar{b}_{m}^{d} b_{h}^{L}(M) \xi ^{L}_{mm}(\textbf{x}) \,, \\
\xi _{vd}(\textbf{x}|M) &=& \bar{b}_{m}^{d} b_{v}^{L}(M) \xi ^{L}_{mm}(\textbf{x}) \,, \\
\xi _{dd}(\textbf{x}|M) &=& \left(\bar{b}_{m}^{d}\right) ^{2} \xi ^{L}_{mm}(\textbf{x}) \,, 
\end{eqnarray}
where $\bar{b}_{m}^{d}$ is the mean bias of the matter in dust, defined as
\begin{equation} \label{eq:dust_bias}
\bar{b}_{m}^{d} \equiv 1 - \bar{b}_{m}^{h} - \bar{b}_{m}^{v} \,,
\end{equation}
to guarantee that the total matter is not biased with respect to itself.
The mean matter density is now given by
\begin{equation} 
\bar{\rho} _{m} = \bar{\rho}^{h} _{m} + \bar{\rho}^{v} _{m} + \bar{\rho}^{d} _{m} \,,
\end{equation}
such that this expression naturally defines the matter density in dust $\bar{\rho}^{d} _{m}$.

The matter power spectrum in the HVDM is given by
\begin{eqnarray}
P_{mm}(k) &=&  P_{mm}^{1H}(k) + P_{mm}^{2H}(k) + P_{mm}^{1V}(k) + P_{mm}^{2V}(k) \nonumber\\  &+&  2P_{mm}^{HV}(k) + 2P_{mm}^{HD}(k) + 2P_{mm}^{VD}(k) + P_{mm}^{2D}(k) \,,
\label{eq:Halo_Void_Dust_Model}
\end{eqnarray}
where new terms with dust are given by
{\small \begin{eqnarray}
P_{mm}^{HD}(k) &=& \bar{b}_{m}^{d} \int d \ln M \frac{M}{\bar{\rho}_{m}} \frac{d n_{h}}{d \ln M} u_{h}(k| M) b^{L}_{h}(M) P _{mm}^{L}(k) \,, \\
P_{mm}^{VD}(k) &=& \bar{b}_{m}^{d} \int d \ln M \frac{M}{\bar{\rho}_{m}} \frac{d n_{v}}{d \ln M} u_{v}(k| M) b^{L}_{v}(M) P _{mm}^{L}(k) \,, \\
P_{mm}^{2D}(k) &=& \left(\bar{b}_{m}^{d}\right)^{2}  P_{mm}^{L} (k) \,.
\label{eq:PHVD_terms}
\end{eqnarray} }

\subsubsection*{Cross-correlations}

The HVDM adds an extra term to the HVM prediction for the halo-matter and void-matter spectra in Eqs.~(\ref{eq:HVM_phm}) and (\ref{eq:HVM_pvm}). These are respectively:
\begin{eqnarray}
    P_{hm}^{HD}(k|M) = \bar{b}_{m}^{d} b_h^{L}(M) P^L_{mm}(k)\,, \label{eq:HVDM_phm}\\
        P_{vm}^{VD}(k|M) = \bar{b}_{m}^{d} b_v^{L}(M) P^L_{mm}(k)\,  \label{eq:HVDM_pvm}. 
\end{eqnarray}
For the cross-correlations, the corrections to Eqs.~(\ref{eq:xi_profile_h}) and (\ref{eq:xi_profile_v}) are
\begin{eqnarray}
    \xi_{hm}^{HD}(\textbf{x}|M) = \bar{b}_{m}^{d} b_h^{L}(M) \xi^L_{mm}(r)\,, \label{eq:xi_profile_h_dust} \\
        \xi_{vm}^{VD}(\textbf{x}|M) = \bar{b}_{m}^{d}  b_v^{L}(M) \xi^L_{mm}(r)\, \label{eq:xi_profile_v_dust}. 
\end{eqnarray}

We now proceed to present and fit the ingredients of the HVM and HVDM. On top of the halo properties (profile, abundance and linear bias), which are required in the HM, we also need the corresponding void properties for the HVM and the HVDM.

\section{Ingredients}
\label{sec:halovoid_ingredients}

In this section we consider the ingredients for both halo and void sectors: the halo/void mass function, density profiles and linear bias. Our goal here is to define each of these ingredients and highlight underlying relevant points for the construction of the HV(D)M. Specially important is the fact that the halo/void mass function and bias used naturally incorporate the exclusion of voids inside halos and halos inside voids, preventing double counting of density terms. Moreover, the constraint in Eq.~(\ref{eq:hvm_sum}) is automatically fulfilled, as will be described in Sec.~\ref{sec:consistence}. 

For the calibration of the free parameters of each ingredient needed by the HV(D)M, we have run a N-body simulation with the \texttt{RAMSES} code \cite{Teyssier} in a box of $512$ Mpc$/h$ and with $512^{3}$ particles using the same cosmology as that of the Multidark simulations \cite{Klypin2}, namely ($\Omega _{m}$, $\Omega _{b}$, $\Omega _{\Lambda}$, $\sigma _{8}$, $n_{s}$, $H_{0}$) = (0.307, 0.048, 0.693, 0.829, 0.96, 67.8).

For constructing the halo and void catalogues we first performed a Delanay triangulation of the particle catalogue using the \texttt{CGAL} library. We then considered the position of each particle as the center of a possible halo and the position of each vertex of the Voronoi tessellation (the graph dual to the Delaunay triangulation) as the center of a possible void. We sorted these centers by their local density, estimated using the triangulation, and grew a sphere around this position until its average contrast density is smaller than $360$ for halos, or larger than $0.2$ for voids. We also stopped growing spheres if they hit another previously found halo or void, in order to avoid double counting of matter. In order to measure the bias and the cross-correlations, we divided halos and voids in eight different bins of masses and radius (see Table~\ref{table:Mass_bins}). 

\begin{table}[t!]
\begin{center}
\begin{tabular}{ | c || c | c || c | c | } 
\hline
\rm{Bin} & $M _{h}$ $[M_{\odot}/h]$ & $N_{h}$ & $r_{v}$ $[{\rm Mpc}/h]$ & $N_{v}$  \\ 
\hline
\hline
0 & $5.25 \times 10^{12}$ & 56414 & 3.32 & 15657 \\
\hline
1 & $1.23 \times 10^{13}$ & 31453 & 4.03 & 11796 \\
\hline
2 & $2.88 \times 10^{13}$ & 15913 & 4.91 & 6693 \\
\hline
3 & $6.74 \times 10^{13}$ & 6574 & 5.99 & 3028 \\
\hline
4 & $1.57 \times 10^{14}$ & 2229 & 7.29 & 1109 \\
\hline
5 & $3.70 \times 10^{14}$ & 557 & 8.93 & 362 \\
\hline
6 & $8.07 \times 10^{14}$ & 88 & 10.89 & 80 \\
\hline
7 & $1.78 \times 10^{15}$ & 4 & 13.07 & 17 \\
\hline
\end{tabular}
\caption{Halo and void bins considered in this work. For each bin, we display the mean mass or radius and the number of objects.} 
\label{table:Mass_bins}
\end{center} 
\end{table}


\subsection{Profile} \label{sec:profiles}

We now describe the halo and void profiles used in this work. For halos, we used the standark Navarro--Frenk--White (NFW) profile \cite{NFW}. For voids, we propose a new profile that is in better agreement with the void finder used here, compared to the typical void profile used in other works \cite{Hamaus2}.

\subsubsection*{Halo}

For halos, the most widely used density profile is the Navarro--Frenk--White (NFW) profile \cite{NFW}
\begin{equation} \label{eq:nfw}
\rho_{\rm NFW}(r | M) = \frac{\rho_s}{c(M)r/r_{\rm vir}(1+c(M)r/r_{\rm vir})^2} \,,  
\end{equation}
where $\rho_s$ is the characteristic density (a normalization), $c$ is the concentration parameter \cite{Bullock} and $r_{\rm vir}$ is the virial radius defined as the radius of a sphere with density equal to $\Delta_{\rm vir}$ times the matter density of the universe. $\Delta_{\rm vir} = 360 $ in this work, which is very close to the virial overdensity computed using spherical collapse ($\Delta _{\rm vir} \approx 334$). The NFW profile describes two power laws -- $r^{-1}$ for  $r\ll r_{\rm vir}/c$ and $r^{-3}$ for  $r \gg r_{\rm vir}/c$. The profile's Fourier transform, truncated at $r_{\rm vir}$, is given by
\begin{equation}
    u_h(k|M) = \int^{r_{\rm vir}}_0 \frac{4\pi r^2}{M}\frac{\sin{kr}}{kr}\rho_{\rm NFW}(r|M)dr\,.
\end{equation}
See also \cite{Scoccimarro:2000gm} for an analytical expression.

We compare the NFW profile with the measurements from our simulation in the left panel of Fig.~\ref{fig:profiles_hv}. We can see that the NFW profile agrees with the data up to twice the halo radius, so we will use it in this work to describe the matter distribution inside halos. The matter distribution in the outer regions is under-predicted because this matter comes from other structures. As we will show in Sec.~\ref{sec:mh_results}, the matter distribution in the outer region is well explained by the HM as well as by the HV(D)M, which add a correction accounting for the matter coming from nearby halos and voids.

\subsubsection*{Void}

For voids, the most widely used density profile is the Hamaus--Sutter--Wandelt (HSW) profile \cite{Hamaus2}
\begin{equation}\label{eq:HSW_profile}
    \frac{\rho_v(r)}{\bar{\rho}_m} - 1 = \delta_c \frac{1-(r/r_s)^\alpha}{1+(r/r_v)^\beta} \,,
\end{equation}
where $r_s$ is the radius for which $\rho_v = \bar{\rho}_m$, $r_v$ is the effective void radius and $(\alpha,\beta,\delta_c )$ are free parameters. Notice that $\delta_c$ in this equation is the void central density, and not the critical density computed in spherical collapse.

\begin{figure}[h]
	\centering
	\includegraphics[width=0.49\textwidth]{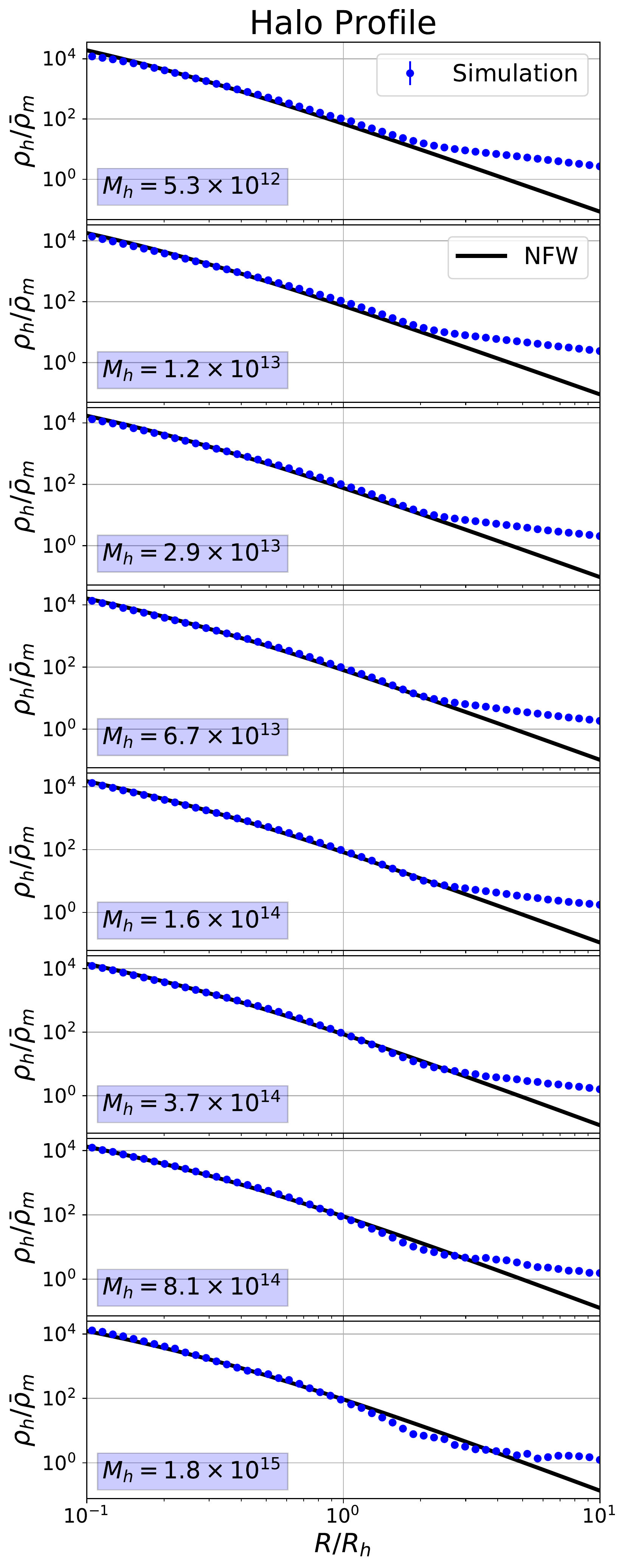}
	\includegraphics[width=0.49\textwidth]{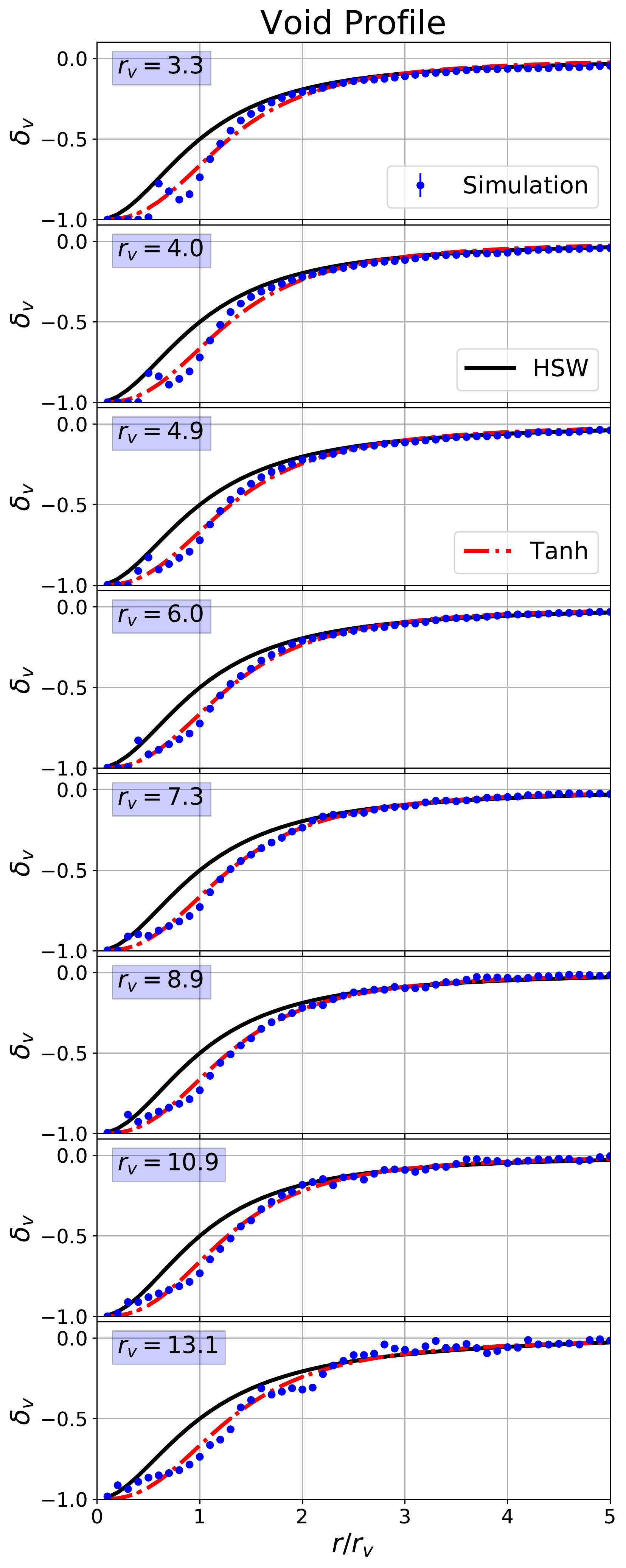}
	\caption{\label{fig:profiles_hv} \small ({\it Left:}) the measured halo profile (blue dots) compared to the NFW prescription (black). ({\it Right:}) the measured void profile $\delta_v=\rho_v/\bar{\rho}_m-1$ compared to the HSW profile \cite{Hamaus2} (black) and the profile proposed in this work (red dashed). Halo masses $M_h$ are given in units of $M_{\odot}/h$ and the void radii $r_v$ are in units of ${\rm Mpc}/h$. 
	}
\end{figure}

Since the void finder used in this work identifies voids with an empty center, as explained above, we propose a new fitting function. We use a hyperbolic tangent with a single free parameter, in such a way that its value and its first derivative vanish in the center 
\begin{equation}\label{eq:void_profile}
\frac{\rho_v(r|r_v)}{\bar{\rho}_m}  = \frac{1}{2} \left[ 1 + \tanh{ \left(\frac{y - y_{0}}{s(r_v)} \right)} \right] \,,
\end{equation}
where $y = \ln{\left( r/r_{v}\right)}$ and $y_{0} = \ln{\left( r_{0}/r_{v}\right)}$. The radius  
$r_{0}$ is fixed by requiring that the profile integral up to $r_v$ is $\Delta_v = 0.2$, such that we can parameterize $r_{0}(s)$ (in units of  Mpc$/h$) as a second order polynomial function: $r_{0}(s) = 0.37s^{2} + 0.25s +  0.89$, where $s$ is the single free parameter for this profile. The parameter $s$ plays a similar role of the concentration parameter in the NFW profile, as it determines how fast the density grows as we move away from the void center. 
For the simulation and specific void finder considered in this work, we have checked that this parameter depends very weakly on the void radius, so we have fixed it to $s = 0.75$, which is the average value of the best fit for all bins of void radius.
Similarly to what was done for halos, we define the Fourier transform of the void profile as 
\begin{equation}
    u_v(k|M) = \int^{r_{v}}_0 \frac{4\pi r^2}{M}\frac{\sin{kr}}{kr}\rho_{v}(r|M)dr \,.
\end{equation}

In the right panel of Fig.~\ref{fig:profiles_hv}, we compare the HSW profile, the $\tanh$ profile and the profile measured from our simulation. We have fitted the three\footnote{As the center of the voids is empty, the value of $\delta _{c}$ is fixed in $-1$.} free parameters of the HSW profile because our void finder is different from that used in \cite{Hamaus2}. Since the void finder used here does not rely on any central particle, but in Voronoi vertices, we will use here the profile in Eq.~\eqref{eq:void_profile}, which is more accurate than the HSW profile. Notice from Fig.~\ref{fig:profiles_hv} that the measured profiles are less dense in the center and do not contain any compensated wall, two features that differ from the measured density profiles presented in \cite{Hamaus2}.

However, the authors would like to highlight that the profile proposed in Eq.~\eqref{eq:void_profile} describes better just the spherical voids like the ones used in this work. Voids found using the widely used void finder \texttt{VIDE} \cite{VIDE} are better described by the void density profile first presented in \cite{Hamaus2}. It is also important to note that the dependence of the void density profile on the specific void finder is an open topic and deserve more attention in future works.

\subsection{Mass function} \label{sec:mass_func}

We now describe void and halo abundances. In this section we present the functional forms of these expressions and leave the computation of excursion set theory based functions for App.~\ref{app:abundance_bias}. We also show the best-fits for each model and compare them to the N-body simulation results.

An important quantity used to compute the mass functions and the linear bias is the variance of the linear density field smoothed at some scale $R$, defined as 
\begin{equation}\label{eq:variance}
    S(R) \equiv \sigma^2(R) \equiv \int \frac{dk}{2\pi^2}k^2P^{L}_{mm}(k)|\tilde{W}(k|R)|^2 \,,
\end{equation}
where $\tilde{W}$ is the window function in Fourier space, taken in this work and usually in the literature as a top-hat function in real space.
Since the excursion set theory computations are performed in Lagrangian space \cite{Bond, Maggiore1}, the relation between radius $R$ and  mass $M$ of each structure is given by
\begin{equation}
     M = \frac{4}{3}\pi R^3 \bar{\rho}_{m}  \,.
\end{equation}

\subsubsection*{Halo}

The halo mass function gives the differential number density of halos in a given bin of mass $M$, and is usually written as 
\begin{equation}
    \frac{dn_h}{d\ln{M}} = f_h(\sigma)\frac{\bar{\rho}_m}{M}\frac{d\ln{\sigma^{-1}}}{d\ln{M}}\,,
\end{equation}
where $f_{h}(\sigma)$ is the multiplicity function and its particular choice determines the model.

In this work, we consider the following multiplicity functions for halos:
\begin{eqnarray}
    f^{\rm 1SB}_h(\sigma) &=& \left( \frac{2}{\pi}\right)^{\frac{1}{2}}\frac{\delta_c}{\sigma}\exp{\left(-2\delta_c^2/2\sigma^2\right)} \, , \\
        f^{\rm Tinker}_h(\sigma) &=& A\left[ \left( \frac{\sigma}{b}\right)^{-a} + 1\right]\exp{\left(-c/\sigma^2 \right)}\,, \\
    f^{\rm 2LDB}_h(\sigma) &=& 2(1+D_h)\exp{\left[-\frac{\beta_h^2\sigma^2}{2(1+D_h)}-\frac{\beta_h\delta_c}{1+D_h}\right]}
   \nonumber \\ &\times& \sum_n \frac{n\pi}{\delta_T^2}\sigma^2\sin \left( \frac{n\pi\delta_c}{\delta_T} \right)\exp{\left[-\frac{n^2\pi^2(1+D_h)}{2\delta_T^2}\sigma^2\right]} \label{eq:massfunc_2ldb}\,,
\end{eqnarray}
where $\delta_c$ is the critical density for halo formation linearly extrapolated ($\delta_c \approx 1.686$ for EdS), $\delta_{v}$ is the critical density for void formation linearly extrapolated  ($\delta_v \approx -2.7$ for EdS), and $\delta_T = |\delta_v| + \delta_c$. Here $f^{\rm 1SB}_{h}$ is the Press--Schechter mass function \cite{Press}, where the label $\rm 1SB$ means one static barrier. Likewise, 
$f^{\rm Tinker}_{h}$ is the Tinker \textit{et al.} mass function an $(A, a, b, c)$ are free parameters \cite{Tinker}. Finally, $f^{\rm 2LDB}_{h}$ is the excursion-set prediction for two linear diffusive barriers (2LDB) and $(\beta _{h}, D_{h})$ are free parameters describing respectively the slope and the diffusive coefficient of the barriers \cite{Voivodic1}. 

It is important to notice a special case of the 2LDB model, in which one takes $\beta _{h} = D_{h} = 0$. This describes two static barriers (2SB) \cite{Sheth1, Jennings}
\begin{equation}
    f^{\rm 2SB}_h(\sigma) = f^{\rm 2LDB}_h(\sigma, \beta_h = 0, D_h = 0) \,.
\end{equation}
For that case, the excursion set realizations for sure will cross one of the two barriers, such that the constraint in Eq.~\eqref{eq:hvm_sum} is naturally satisfied (more about this in Sec.~\ref{sec:consistence} and App.~\ref{app:abundance_bias}). 

We note that the $f^{\rm 1SB}_{h}$ and $f^{\rm Tinker}_{h}$ mass functions are normalized to unity whereas $f^{\rm 2LDB}_{h}$ is not, because it already takes into account the existence of voids. In fact, $f^{\rm 2LDB}_{h}$ will only be properly normalized when we also consider the contribution coming from voids and correctly choose free parameters $(\beta, D)$ (more about this in Sec.~\ref{sec:consistence}). 

In the left panel of Fig.~\ref{fig:abundances}, we compare the measured abundance and predictions of the models above. We see that both the Tinker and the 2LDB models are in good agreement with the simulation with a scatter $\sim 10\%$ while the PS model has a larger deviation of $\sim 40\%$ for small masses. The fitted values of $D_h$ and $\beta_h$ are shown in Table~\ref{table:params1}. 
We do not display results for the 2SB case because its mass function provides very similar results to the 1SB case in the mass range displayed in the plots; in other words, the cloud-in-void effect is negligible in this range.

\subsubsection*{Void}
For cosmic voids, the mass function is usually expressed in terms of the void radius $R$
\begin{equation}
    \frac{dn_v}{d\ln{R}} = \frac{f_v(\sigma)}{V(R)}\frac{d\ln{\sigma^{-1}}}{d\ln{R}}\,,
\end{equation}

Another difference is that the quantity conserved from the linear to the non-linear theory is not the number density, as is the case for halos. As discussed in Jennings \textit{et al.} \cite{Jennings}, the correct quantity to be conserved is the volume density of voids
\begin{equation}
    V(r) dn = V(r_{L}) dn_{L} |_{r_{L}(r)} \,,
\end{equation}
where the relation between the linear and the non-linear radius of voids is $r = \Delta _{v} ^{-1/3} r_{L} \approx 1.71 r_{L}$.

In this work, we consider the following multiplicity functions for voids:
\begin{eqnarray}
    f^{\rm 1SB}_v(\sigma) &=& \left( \frac{2}{\pi}\right)^{\frac{1}{2}}\frac{\delta_v}{\sigma}\exp{\left(-2\delta_v^2/2\sigma^2\right)} \,, \\
    f_v^{\rm 2LDB}(\sigma) &=&  2(1+D_v)\exp{\left[-\frac{\beta_v^2\sigma^2}{2(1+D_v)}-\frac{\beta_v|\delta_v|}{1+D_v}\right]}
    \nonumber \\ &\times& \sum_n \frac{n\pi}{\delta_T^2}\sigma^2\sin \left( \frac{n\pi |\delta_v|}{\delta_T} \right)\exp{\left[-\frac{n^2\pi^2(1+D_v)}{2\delta_T^2}\sigma^2\right]}\,, \label{eq:massfunc_2ldb_void}
\end{eqnarray}
where $\delta _{c}$, $\delta_{v}$ and $\delta _{T}$ are the same quantities presented in the halo subsection. $f^{\rm 2LDB}_{v}$ is the excursion set prediction for two linear diffusive barrier with $(\beta_{v}, D_{v})$ being free parameters describing the slope and the diffusion coefficient of the barriers \cite{Voivodic1}. Again, $f^{\rm 1SB}_{v}$ stands for Press--Schechter prediction, which has a single static barrier. We also have for voids the double static barrier (2SB) as a limiting case of 2LDB:
\begin{equation}
    f^{\rm 2SB}_v(\sigma) = f^{\rm 2LDB}_v(\sigma, \beta_v = 0, D_v = 0) \,.
\end{equation}

\begin{figure}[ht]
\centering
  \includegraphics[width=0.49\textwidth]{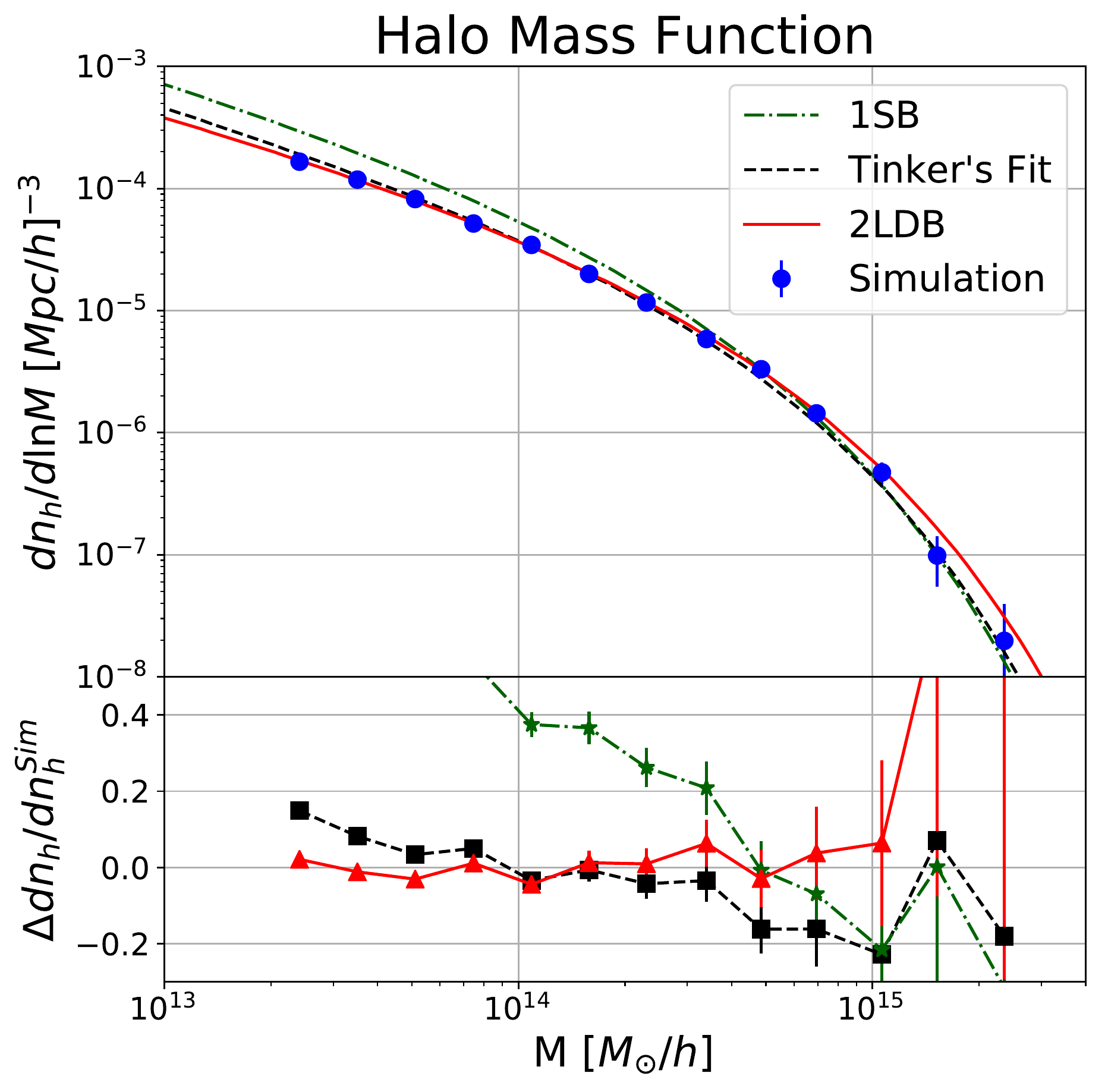}
  \includegraphics[width=0.49\textwidth]{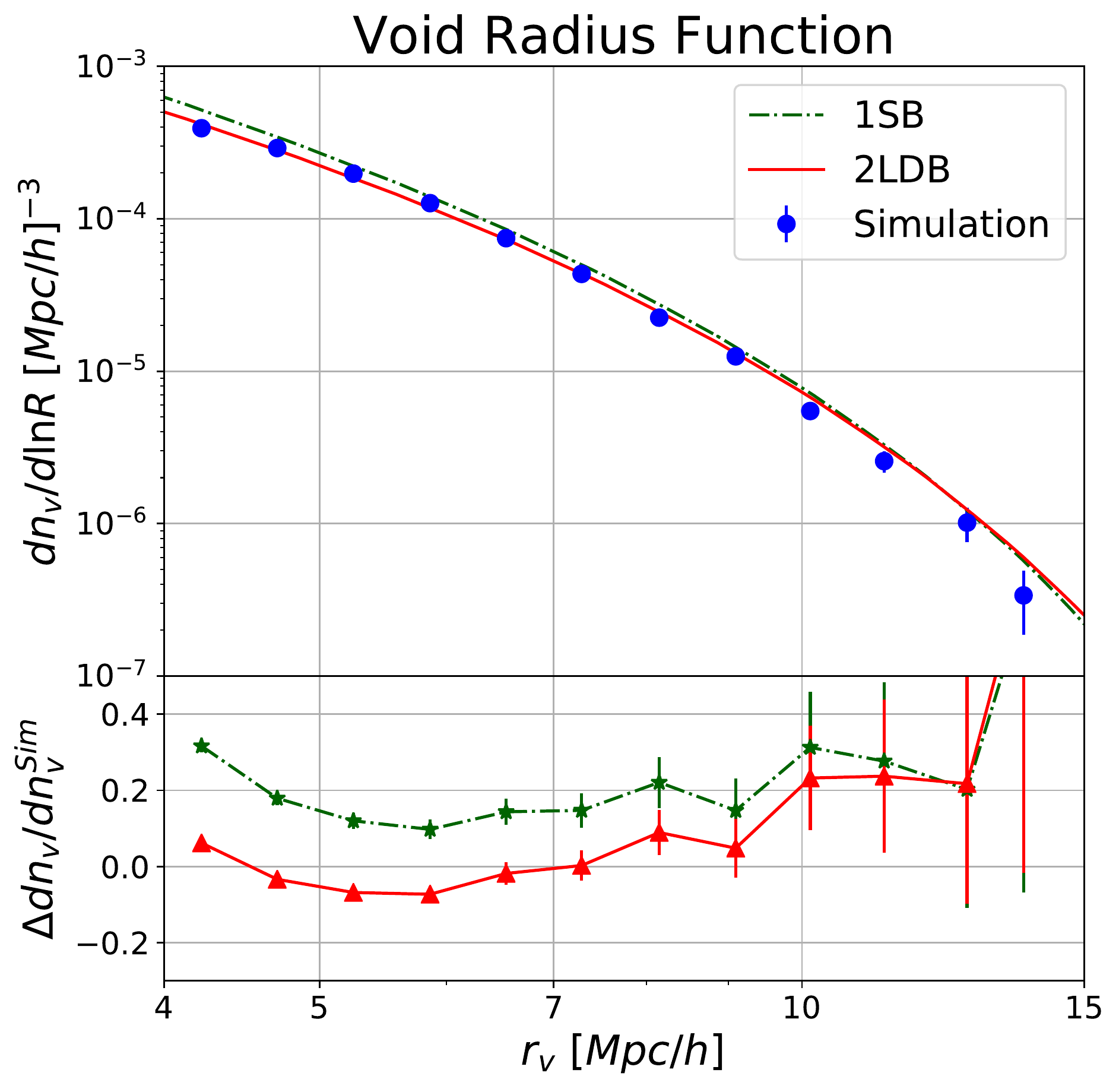} 
\caption{\label{fig:abundances}%
\small Abundance function of halos as a function of halo mass $M$ ({\it left}) and voids as a function of void radius $r_v$ ({\it right}). The bottom panels show the relative difference of the models with respect to mass-function measure in our simulation.  We compare three different models: one static barrier (1SB) or Press--Schechter (dot-dashed green line), Tinker (dashed line) and two linear diffusive barriers (2LDB, solid red line). For both halos and voids, the 2LDB model displays the best overall results. }
\end{figure}

It turns out that the 2SB case gives very similar results to the 1SB case in the range void radii considered here. In the right panel of Fig.~\ref{fig:abundances}, we compare the measured void abundance with the prediction of the models above. The values of $D_v$ and $\beta_v$ for void abundance are in Table~\ref{table:params1}. We see that both models slightly overestimate the abundance for large radii and that 2LDB describes better the abundance for radii smaller than $10$ Mpc$/h$, where the measurements have smaller error bars.

\subsection{Linear bias and the power spectrum exclusion term} \label{sec:bias}

Besides the density profile and abundance, other fundamental ingredients for the HV(D)M are the halo-halo, void-void and halo-void power spectra (see Eqs.~(\ref{eq:P2halo}), (\ref{eq:P2void}) and (\ref{eq:Phv}) ). In order to express $P_{hh}$, $P_{vv}$ and $P_{hv}$ in terms of the linear matter-matter power spectrum with some dependence on halo mass (or void radius), we can write in the most general way at the level of the density fields
\begin{equation}
\delta_{x}(k|M) = \mathcal{F}_{x}[\delta_{m}^{L}(k), k, M] \,,
\label{eq:Pxy}
\end{equation}
where $x = $ halo ($h$) or void ($v$).
The map in Eq.~\eqref{eq:Pxy} can be computed accurately, up to shell-crossing, using perturbation theory \cite{Assassi}. In this work we use only linear approximation (tree-level)
\begin{equation}
    \delta_x(k,M) = b_x^{L}(M)\delta_m(k) \,,
\end{equation}
where $b_{x}^{L}(M)$ is the linear bias of  structure $x$ with mass $M$. From now on, we drop the $L$ label on $b$, and write
\begin{eqnarray}
P_{hh}(k|M_{1}, M_{2}) &=& b_h(M_1)b_h(M_2)P_{mm}^L(k) \,,  \\ 
P_{vv}(k|M_{1}, M_{2}) &=& b_v(M_1)b_v(M_2)P_{mm}^L(k) \,,  \\
P_{hv}(k|M_{1}, M_{2}) &=& b_h(M_1)b_v(M_2)P_{mm}^L(k) \,. 
\end{eqnarray}

Besides the tree level expansion of the $xy$ power spectra, we will also consider the exclusion term implemented by \cite{Baldauf} for halos and by \cite{Chan} for voids. This term suppresses the correlation function on scales smaller than the sum of the radii of the two structures and will be specially relevant in the case of the void-matter correlations in Sec.~\ref{sec:mv_results}. The exclusion term is crucial to suppress the 2Void and the Halo-Void contribution in the innermost part of the void. At tree level and taking into account the exclusion term, the $xy$ correlation is
\begin{equation}
\xi _{xy}(r|M_{1}, M_{2}) = W(r|D_{xy}) \left[ b_{x}(M_{1})b_{y}(M_{2})\xi _{mm}(r) + 1\right] - 1 \,,
\label{eq:Xixy}
\end{equation}
where $W(r|D_{xy})$ is the exclusion term between two structures with separation $D_{xy} = R_{x} + R_{y}$, which must vanish for $r \lesssim D$ and must be unity for $r \gtrsim D$.

Therefore, the $P_{xy}$ power spectrum is given by
\begin{eqnarray}
P_{xy}(k|M_{1}, M_{2}) &=& 4 \pi b_{x}(M_{1})b_{y}(M_{2}) \int _{0} ^{\infty} dr \, r^{2} \, \xi _{mm}^{L}(r)W(r|D_{xy}) j_{0}(kr) \nonumber \\ 
&+& 4\pi \int _{0} ^{\infty} dr \, r^{2} \, \left[ W(r|D_{xy}) - 1 \right] j_{0}(kr) \,,
\end{eqnarray}
where $j_{0}(x)$ is the spherical Bessel function of zeroth order.
Typical choices for $W(r|D)$ are the top-hat function or 
the hyperbolic tangent ($\tanh$) function
\begin{equation}
W_{\rm Tanh} (r|D) = \frac{1}{2} \left[ 1 + \tanh \left( \frac{\ln r - \ln D}{\tilde{\sigma}} \right)\right] \,,
\end{equation}
where $\tilde{\sigma }$ is a free parameter that controls the transition between $W(r|D)=0$ and $W(r|D)=1$, and was fixed to $\tilde{\sigma} = 0.1$ in this work.
 
In principle $\tilde{\sigma}$ can also be changed but the interpretations of our results should not depend on its particular value. In fact the value of $\tilde{\sigma}$ is degenerate with the choice of the contrast overdensity (or radius) for defining halos and voids, since $\tilde{\sigma}$ controls how much matter outside the structure will be taken into account. For simplicity, in this work we will not consider different values for either 
$\tilde{\sigma}$ or the contrast overdensity.

We now proceed to the functional dependence of the bias on the rms $\sigma(R) = \sqrt{S(R)}$ of the density field for halos and voids. We propose a new bias that comes from the excursion set theory \cite{deSimone2} using two linear diffusive barriers ($b^{\rm 2LDB}$) and compare it to our simulation results. Notice that using this bias makes our framework fully self-consistent with the 2LDB abundance (see Eqs.~\eqref{eq:massfunc_2ldb} and \eqref{eq:massfunc_2ldb_void}). Furthermore, using both 2LDB abundance and 2LDB bias (for both halos and voids) is vital to exclude the overlap of structures and avoid double-counting of matter. The theoretical derivation of the 2LDB bias is presented in App.~\ref{app:abundance_bias}.

\subsubsection*{Halo}

The linear halo bias expressions considered in this work are listed below:
\begin{eqnarray}
b_{h}^{\rm 1SB}(\sigma) &=& 1 + \frac{\nu^2-1}{\delta_c}\,, \\
b_{h}^{\rm Tinker}(\sigma) &=& 1 - A \frac{\nu^a}{\nu^a+\delta_c^a} + B\nu^b + C\nu^c \,,\\
b_{h}^{\rm 2LDB}(\sigma) &=& 1-  \frac{\sum_n \frac{n\pi}{\delta_T^2}\sin \left( \frac{n\pi \delta_c}{\delta_T}\right)\exp \left[-\frac{n^2\pi^2(1+D_h)}{2\delta_T^2}\sigma^2\right]\left[ \mbox{cotan} \left( \frac{n\pi \delta_{c}}{\delta _{T}}\right)\frac{n\pi}{\delta _{T}} - \frac{\beta_h}{1 + D_h} \right]} {\sum_n \frac{n\pi}{\delta_T^2}\sin \left( \frac{n\pi \delta_c}{\delta_T}\right)\exp \left[-\frac{n^2\pi^2(1+D_h)}{2\delta_T^2}\sigma^2\right] } \label{b_halo} \,,
\end{eqnarray}
where $\nu = \delta_c/\sigma$, $b_{h}^{\rm 1SB}$ is the single static barrier (or Press-Schechter) bias \cite{Press} and $b^{\rm Tinker}_{h}$ is the Tinker \textit{et al.} bias, where $(A, B, C, a, c)$ are the same free parameters from \cite{Tinker2}. The bias $b^{\rm 2LDB}_{h}$ corresponds to the two linear diffusive barriers model. This is the linear bias derived from the same probability density distribution for the first barrier crossing used to derive $f^{\rm 2LDB}_{h}$ (see App.~\ref{app:abundance_bias}). The fits of 2LDB parameters were done using the halo-matter power spectrum, since it is less prone to shot noise. We computed the halo bias in our simulation as the constant term of a linear fit of $P_{hm}(k)/P_{mm}(k)$ using large-scale modes ($k < 0.1$  $h/{\rm Mpc}$).

Notice that since $(\beta _{h}, D_{h})$ are the same free parameters of the mass function, one can fit them using both bias and mass function together. This guarantees that there is no double-counting of structures (e.g. halos in voids)\footnote{See App.~\ref{app:abundance_bias} for more on this.}. Since in this section our main goal is to compare different models for the halo and void linear bias, we fitted the bias separately from the mass function. See the parameter values in Table~\ref{table:params1}.

In the left panel of Fig.~\ref{fig:bias}, we compare the linear halo bias for the models above to those measured from the halo-halo and halo-matter power spectra in our simulation. We can see that all three predictions agree better than $10\%$ with data. The Tinker and 2LDB predictions are consistent with data within $1\sigma$.

\subsubsection*{Void}

For the void linear bias we considered the following models:
{\small
\begin{eqnarray}
b_{v}^{\rm 1SB}(\sigma) &=& 1 + \frac{\delta_{v}}{\sigma ^{2}} - \frac{1}{\delta_v}\,, \\
b_v^{\rm 2LDB}(\sigma) &=& 1 + \frac{\sum_n \frac{n\pi}{\delta_T^2}\sin \left( \frac{n\pi |\delta_v|}{\delta_T}\right)\exp \left[-\frac{n^2\pi^2(1+D_v)}{2\delta_T^2}\sigma^2\right]\left[ \mbox{cotan}  \left( \frac{n\pi |\delta_{v}|}{\delta _{T}}\right)\frac{n\pi}{\delta _{T}} - \frac{\beta_v}{1 + D_v} \right]} {\sum_n \frac{n\pi}{\delta_T^2}\sin \left( \frac{n\pi |\delta_v|}{\delta_T}\right)\exp \left[-\frac{n^2\pi^2(1+D_v)}{2\delta_T^2}\sigma^2\right] }\,,\label{b_void}
\end{eqnarray}}
where $b^{\rm 1SB}_{v}$ is the linear bias for a single static barrier and $b^{\rm 2LDB}_{v}$ is the linear bias for two linear and diffusive barriers. In $b^{\rm 2LDB}_{v}$, $(\beta_{v}, D_{v})$ are the same free parameters of the void abundance. As for the halos, we fitted abundance and bias separately in order to compare these models (see Table~\ref{table:params1}). 

\begin{figure}[ht]
\centering
  \includegraphics[width=0.49\textwidth]{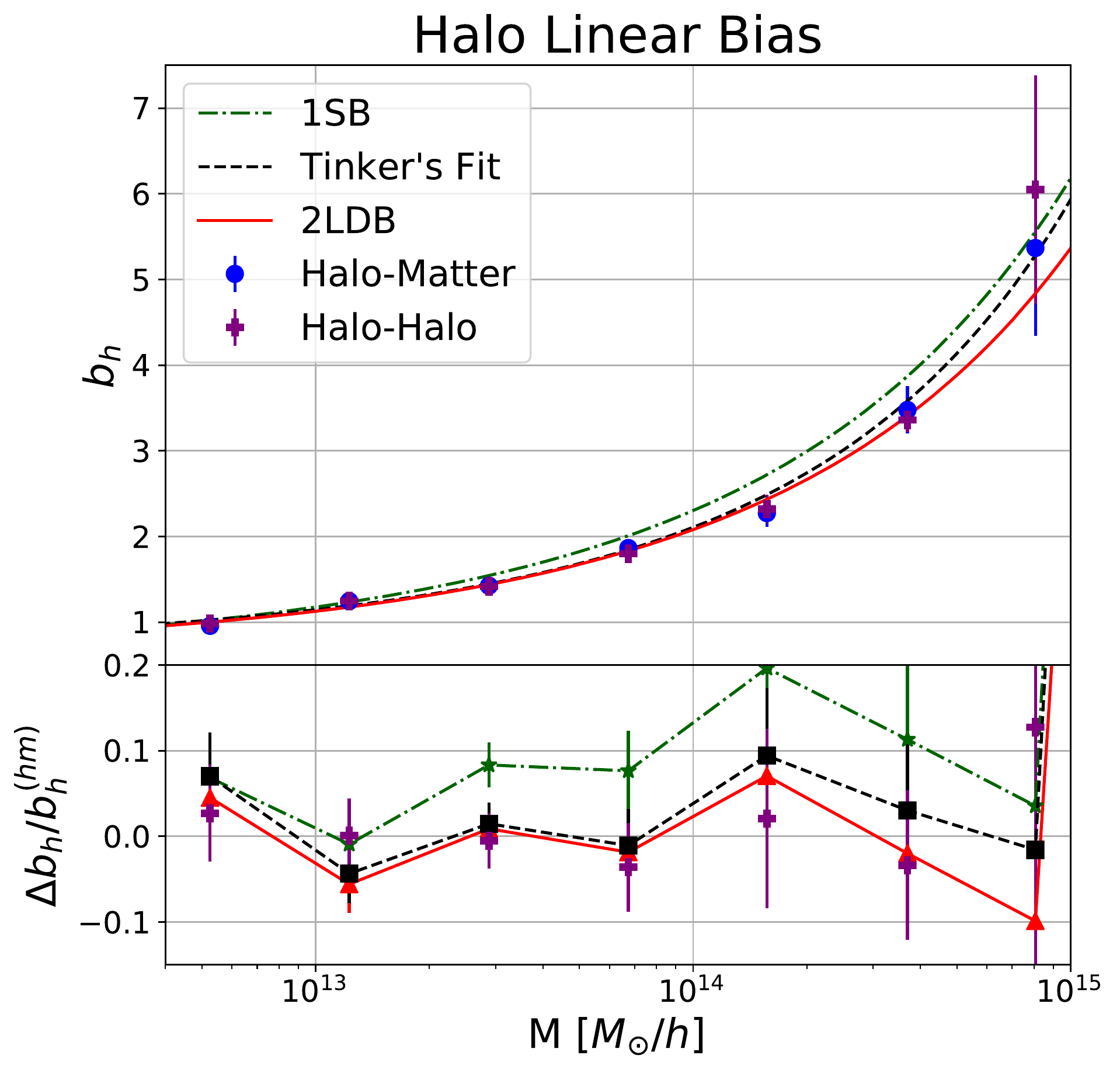}
  \includegraphics[width=0.49\textwidth]{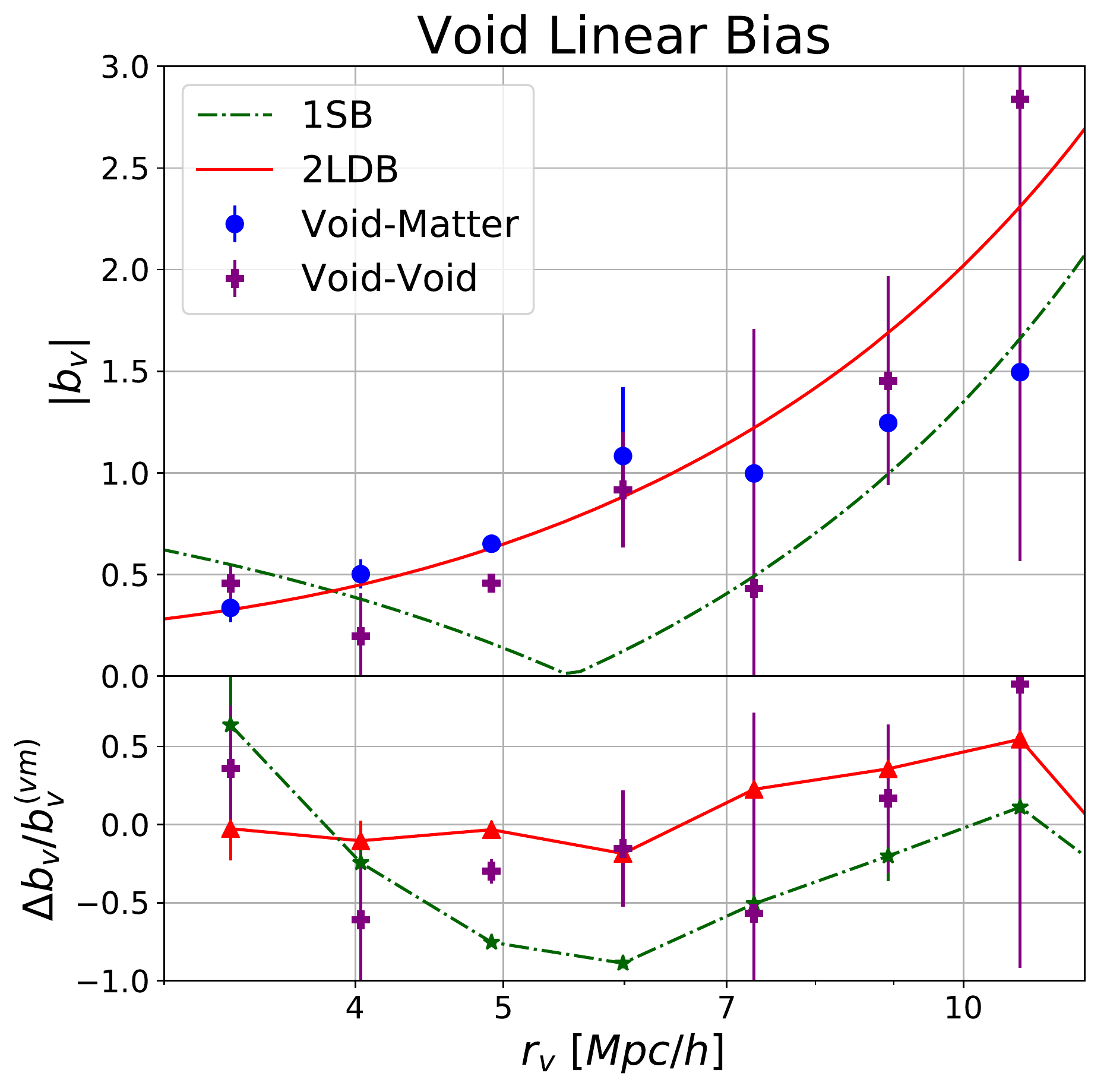} 
\caption{\label{fig:bias}%
\small Linear bias of halos as a function of the halo mass ({\it left}) and bias of voids ({\it right}) as a function of the void radius. The bias is computed in simulations using either the auto-spectra (halo-halo, void-void, circle) or the cross-spectra (halo-matter or void-matter), and compared to the same theory predictions from Fig.~\ref{fig:abundances}. The bottom panels show the relative difference with respect to the bias estimated in simulations via the cross-spectra (halo-matter and void-matter).
Again, the 2LDB model displays the best overall results for both halos and voids.}
\end{figure}

We compare these two linear void bias models to the simulation results (void-void and void-matter power spectra) in the right panel of Fig.~\ref{fig:bias}. We can see that, whereas the 1SB bias does not agree with simulation results, the 2LDB models improves significantly the theoretical predictions, and is consistent with simulations within $1\sigma$ errors. Notice that the shot-noise errors for the void bias are in general larger than for halos, since the number of objects is smaller.

Notice also that in the right panel of Fig.~\ref{fig:bias} we display the absolute value of the linear bias, because the voids used in this work are in general anti-correlated with the matter field. 
This feature is different from what is observed in non-spherical voids \cite{Chan}, but is consistent with the lack of a compensated wall in the void profiles  (see  Fig.~\ref{fig:profiles_hv}). The existence of this compensated wall is determined by the sign of the void linear bias, and is present only when the bias is positive (usually the case for small voids).

\begin{table}[t!]
\begin{center}
\begin{tabular}{ | c || c | c |} 
\hline
\rm{Parameter} & \rm{Abundance} & \rm{Bias} \\ 
\hline
$\beta_h$  & 0.38 & 0.08  \\ 
\hline
$D_h$  & 0.23 & 0.18 \\ 
\hline
$\beta_v$ & 0.09 & 0.92 \\ 
\hline
$D_v$ & 0.06 & 0.07 \\ 
\hline 
\end{tabular}
\caption{ Best-fit values for the two linear and diffusive barriers model (2LDB) parameters. Abundance and bias were fitted separately, without imposing any constraint over the sum of halos and void contributions (see Eq.~\eqref{eq:hvm_sum}). In the case of the HVM results presented in Section~\ref{sec:results}, we have no free parameters since we use the 1SB prescription. For the HVDM, we fit the abundance and bias separately. 
}
\label{table:params1}
\end{center} 
\end{table}

It is worth to point out that the different values found in Table~\ref{table:params1} between the abundances and biases indicate an inconsistency in the modelling. This probably occurs due to the simplified barriers considered in this work (discussed in App.~\ref{app:abundance_bias}) which, besides the good results for the abundance and bias, are an approximation for the non-spherical collapse (expansion) of the halos (voids). In our case, this inconsistency is not important in practice for the results shown in Sec.~\ref{sec:results} because in the case of the HVM we are using no free parameters, but the 1SB prescription. For the HVDM, adding the dust component (by construction) properly normalizes the large scales, independently of the abundances and biases used for the halos and voids.

Moreover, in Sec.~\ref{sec:mv_results}, we show that the best fit for the void bias presented in Table~\ref{table:params1} is consistent with the void profiles measured from the same simulation, while the first is negative for all void bins the second does not have the compensated wall in any bin. This relation between the linear bias and the compensated wall is made clear by Eq.~\eqref{eq:VM_pvm}.

\subsection{Solving the convergence problem} \label{sec:consistence}

We now proceed to demonstrate the consistency of using two barriers for the HVM ingredients. We show that considering halos and voids together speeds up the convergence of the large-scale integrals in Eq.~\eqref{eq:hb_constraint} and Eq.~\eqref{eq:vb_constraint}, indicating that the HVM is a better effective model. 

Considering the 2Halo term integral in Eq.~(\ref{eq:2haloterm}) in the HM, an important constraint is that following integrals of the mass function and bias are properly normalized to unity
\begin{eqnarray}
    I \equiv & \int_{0}^{\infty} d\ln{M} \frac{M }{\bar{\rho}_{m}}   \frac{dn}{d\ln{M}} = 1 \,, \label{eq:I} \\
    I^b \equiv & \int_{0}^{\infty} d\ln{M} \frac{M }{\bar{\rho}_{m}}  \frac{dn}{d\ln{M}}b(M) = 1\,. \label{eq:I_b}
\end{eqnarray}
If this is true, the HM properly recovers the linear matter power on large scales. In order to fulfill these HM constraints, one typically needs either to artificially normalize the 2Halo term or to integrate it down to very small halo masses. The second approach also demands extrapolating the mass function and linear bias expressions to unrealistic low masses, where there is no guarantee of their validity\footnote{Since the abundances approache zero exponentialy fast at large masses for both halos and voids (see Fig.~\ref{fig:abundances}), there is no problem in the upper limit of the integrals in Eqs.~(\ref{eq:I}) and (\ref{eq:I_b}).}.

\begin{figure}[ht]
\centering
  \includegraphics[width=0.49\textwidth]{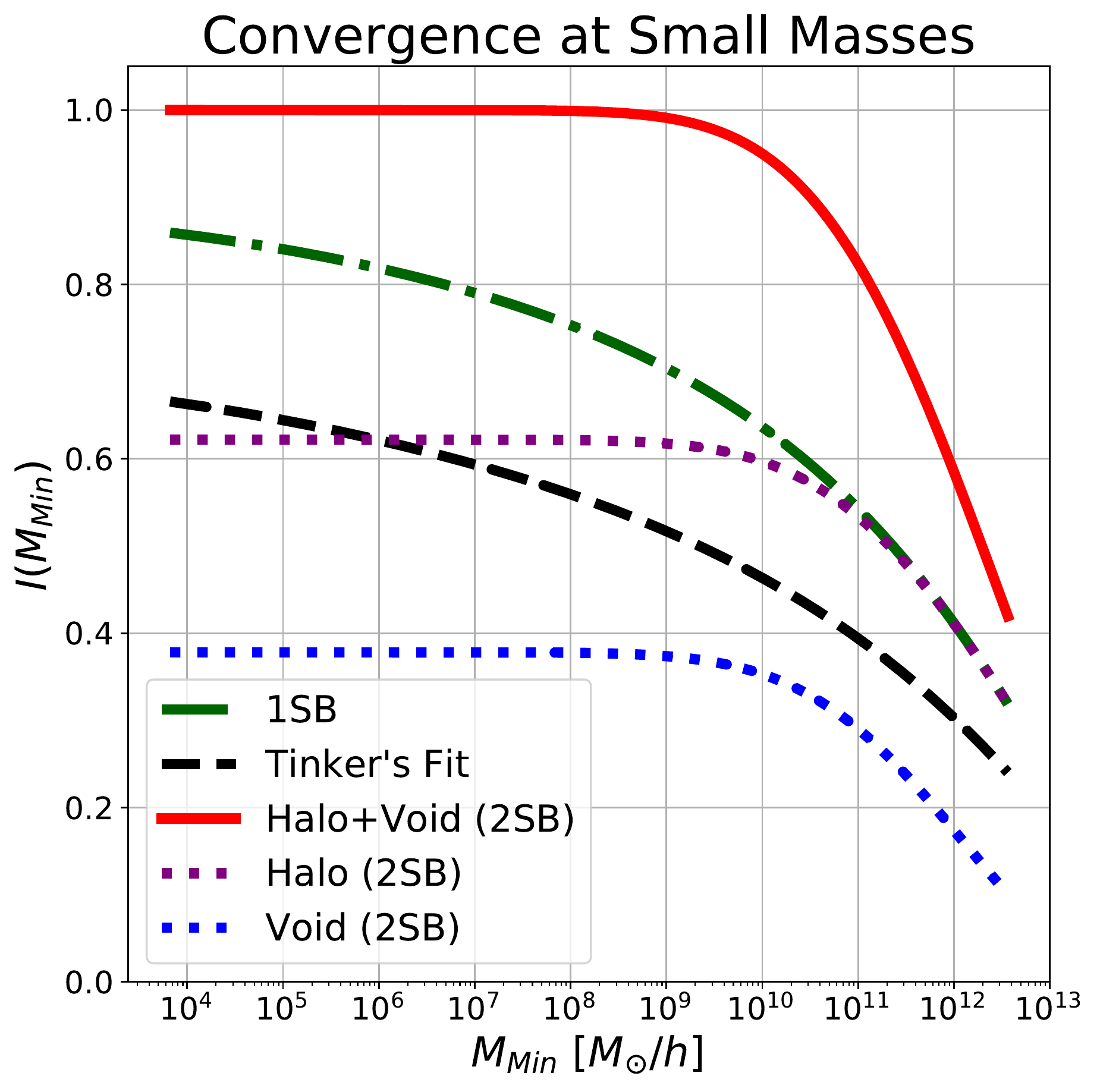}
  \includegraphics[width=0.49\textwidth]{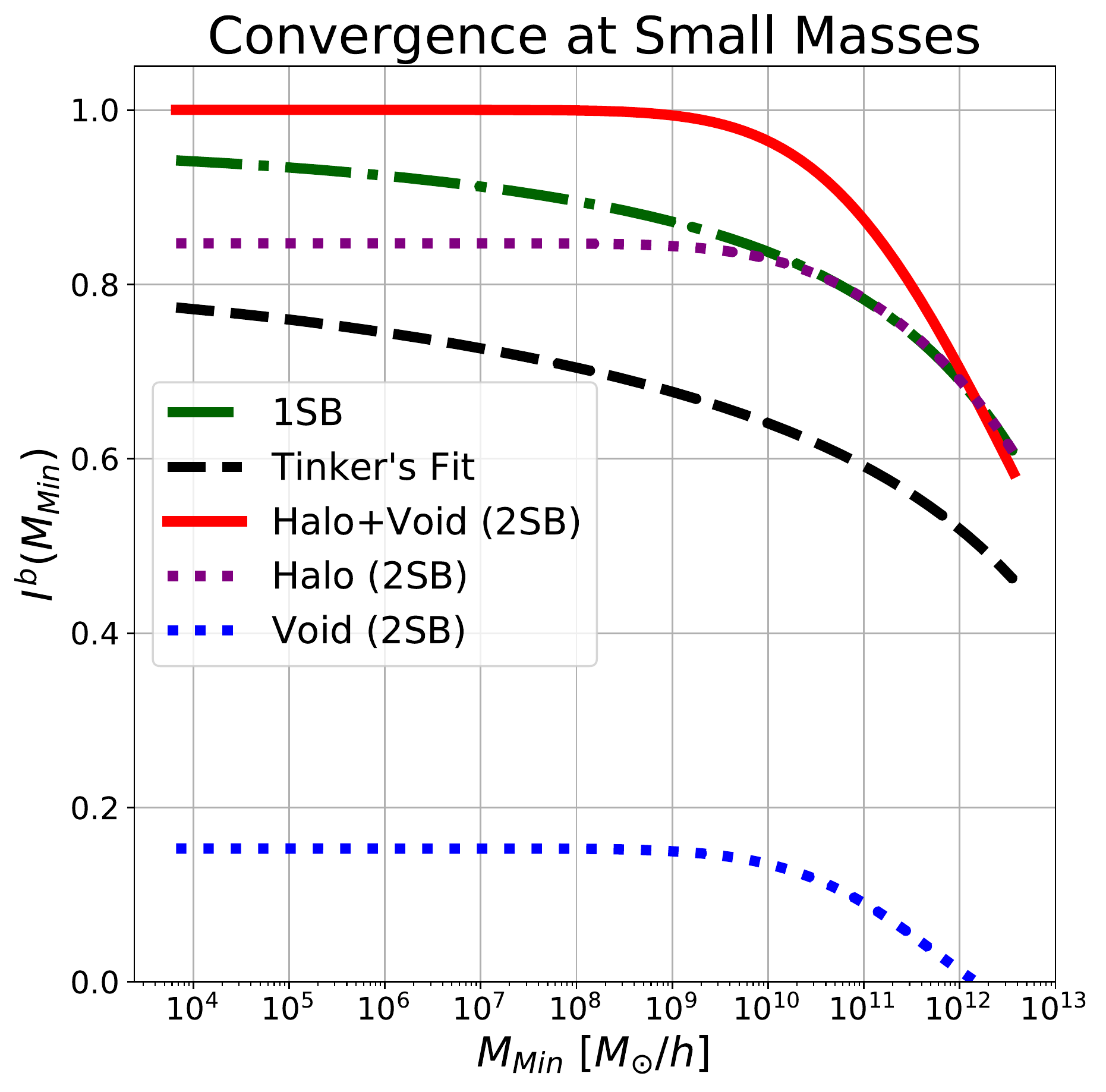} 
\caption{\label{fig:abundances_min} Integral convergence tests. ({\it Left}): We show results of the integral in Eq.~(\ref{eq:I_bot}) for multiple cases. Whereas 1SB (Press--Schechter) and Tinker mass functions did not yet achieved convergence down to $M_{\rm Min} = 10^{4}$ $M_{\odot}/h$, the Halo Void mass function, which considers two barriers, converges at $M_{\rm Min} = 10^{9}$ $M_{\odot}/h$. Notice that the sum of halo and void contribution converges to unit, showing that they naturally satisfy the constraint in Eq.~(\ref{eq:hvm_sum}). ({\it Right}): We show results for the integral in Eq.~(\ref{eq:I_bot_bias}), which displays similar convergence results.}
\end{figure}

Let us now consider how both integrals above depend on their lower limit
\begin{eqnarray}
    I(M_{\rm Min}) &=& \int_{M_{\rm Min}}^{\infty} d\ln{M} \frac{M }{\bar{\rho}_m}   \frac{dn}{d\ln{M}} \label{eq:I_bot}\,, \\
    I^b(M_{\rm Min}) &=& \int_{M_{\rm Min}}^{\infty} d\ln{M} \frac{M }{\bar{\rho}_m}   \frac{dn}{d\ln{M}}b(M) \label{eq:I_bot_bias}\,.
\end{eqnarray}
We display $I$ and $I^b$ respectively in left and right panels of Fig.~\ref{fig:abundances_min}. We consider three models described in the sections above: one static barrier (1SB or Press--Schechter), Tinker and two static barriers (2SB or 2LDB with $\beta_h = D_h = \beta_v = D_v = 0$). As mentioned in Sec.~\ref{sec:mass_func}, this latter case naturally satisfies the constraint in Eq.~(\ref{eq:hvm_sum})\footnote{\label{foot:betachange}When we change the value of $\beta$, the integrals also converge fast, but it might be to a value different from one. That happens because some mass patches might not have enough density to collapse at any scale, since the barrier grows as the mass scale decreases (see Fig.~\ref{fig:Trajectories}). Therefore, in order to keep only halos and voids in the theory, we must scan the $(\beta_h, D_h,  \beta_v,  D_v)$ parameter space imposing the constraint that $\bar{b}_{m}^{d}$ defined by Eq.~(\ref{eq:dust_bias}) is zero. Having values different from zero for $(\beta_h, D_h,  \beta_v,  D_v)$ in such a way that $\bar{b}^{d}_{m} \neq 0$ is consistent with the Halo Void Dust Model, in which the missing matter in the HVM will be taken into account as dust.}.
We see that for both panels of Fig.~\ref{fig:abundances_min}, the integrals for the 1SB and Tinker models did not converge, even as we go to masses lower than $10^{4}$ $M_{\odot}/h$. On the other hand, the integrals have converged for masses of about $10^{9}$ $M_{\odot}/h$ when both halos and voids are considered through the 2SB model.

The faster convergence of the mass function integrals is a key property of the HVM, making clear one of the main features of the model: it counts the mass of smaller halos ($<10^{9}$ $M_{\odot}/h$) as voids of a larger size that are better resolved by the simulations. In that sense, a model with halos and voids is a better effective description of large scales, since it incorporates all UV physics (small scales) into voids well described by the IR physics. 

Note that the Appendix A of \cite{Schmidt2} presents another idea to solve the convergence problem of the large-scale integrals. In that proposal, the integral lower limit is set to a non-zero value and extra terms are added into the mass function, linear bias and stochastic term  to keep mass conservation. Within the formalism developed in the present work, it is straightforward to perform something similar, as we only need to consider halos and dust, i.e. a Halo Dust Model (HDM), in which case the missing matter from small halos will be taken into account as the dust component. The HDM, in that case, would be equivalent to the original Halo Model since the 2Halo term is degenerated with the linear power spectra on large scales, which characterize the dust component. The difference in that case would be the physical meaning taken for the normalization of the matter bias.

\section{Two-point statistics}
\label{sec:results}

Now that we gathered all the ingredients needed for the HVM, we proceed to compare the predictions of the model to our simulation results. In Sec.~\ref{sec:mm_results} we will compare the final matter-matter power spectrum of both HVM and HVDM to the HM prediction. We then proceed to calculate corrections to the cross-spectra and observed density profile of halos in Sec.~\ref{sec:mh_results} and of voids in Sec.~\ref{sec:mv_results}.

Before moving on, let us recall the set up for both HVM and HVDM ingredients\footnote{We emphasize, again, that one reason for the success of the HM is that  it  is  not  a  mere  fit  of  free  parameters  for  the  matter-matter  spectrum  to  simulations. Instead, the HM represents a self-consistent construction over independent halo observables (bias, profile
and mass function), which as a consequence makes predictions for the underlying matter field statistics.}. For the HVM consistency, we must satisfy the constraint in Eq.~(\ref{eq:hvm_constraint}). 
Among the prescriptions given in Sec.~\ref{sec:halovoid_ingredients}, for the HVM we chose to use the 2LDB bias and abundance with $\beta_h = D_h = \beta_v = D_v = 0$, which we label 2SB and naturally satisfies the constraint (see Fig~\ref{fig:abundances_min} and footnote~\ref{foot:betachange})\footnote{In principle one could also explore this four-dimensional parameter space imposing the constraint in Eq.~(\ref{eq:hvm_constraint}) to construct the HVM but it does not improve the results presented here.}. Notice that in the case of the 2SB abundance and linear bias the constraint Eq.~(\ref{eq:hvm_sum}) is automatically fulfilled, as shown in the left panel of Fig.~\ref{fig:abundances_min}.
 The fractions of matter in voids and in halos for this case are 
\begin{equation} 
\frac{\bar{\rho}^{v}_{m}}{\bar{\rho}_{m}} = 0.38\,, \qquad  \frac{\bar{\rho}^{h}_{m}}{\bar{\rho}_{m} } =  0.62\,.
\end{equation}

For the HVDM, we can relax the HVM constraint allowing $\bar{b}_{m}^{d} \neq 0$. In this case, in order to produce the best results for both mass function and bias, we fitted $( \beta_h, D_h, \beta_v, D_v)$ separately. The values obtained for the 2LDB parameters are those described in Table~\ref{table:params1}. In the following we will always use the best fit found using the abundances in the abundances and the best fit found using the linear bias in the linear bias. This space of eight free parameters could be reduced to four free parameters in case we fit the parameters using both the abundance and the linear bias but it does not lead to acceptable results for the voids, since the abundance would dominate the fit jeopardizing the bias. Within the 2LDB set up, we obtained for the fractions of matter in voids, halos and dust
\begin{equation} 
 \frac{\bar{\rho}^{v}_{m}}{\bar{\rho}_{m}} = 0.29\,, \qquad  \frac{\bar{\rho}^{h}_{m}}{\bar{\rho}_{m} } = 0.31\,, \qquad
 \frac{\bar{\rho}^{d}_{m}}{\bar{\rho}_{m} } = 0.4 \,.
\end{equation}

In principle, we could use different prescriptions for the ingredients of both the HVM and the HVDM, which would lead to more precise predictions to some of the correlations. For instance using the Sheth \& Tormen halo mass function \cite{Sheth2} and bias \cite{ST} improves the matter-matter statistics compared to the choices made above. In fact, the HM is very sensitive not only to its ingredients, but also to the choice of radius or overdensity used to truncate the halo profile. For instance, using the Tinker mass-function \cite{Tinker} and bias \cite{Tinker2} with overdensity $\Delta=200$ also leads to significantly better results compared to the virial radius. However, our goal in this work has not been trying to find the prescription that leads to the best overall predictions. Instead, we have aimed to construct a fully {\it self-consistent} theoretical framework, using the spherical collapse and excursion set theory, in which we can compare predictions from the HM and the HV(D)M. In this context, we chose the ingredient prescriptions described above, as well as the virial radius for truncating the halo inner profile. Under our framework, one could make other choices for the ingredient prescriptions in order to get better absolute results for the correlation functions both in the HM and in the HV(D)M. 

\subsection{Matter-matter power spectrum} \label{sec:mm_results}

\begin{figure}[ht]
\centering
\includegraphics[width=0.49\textwidth]{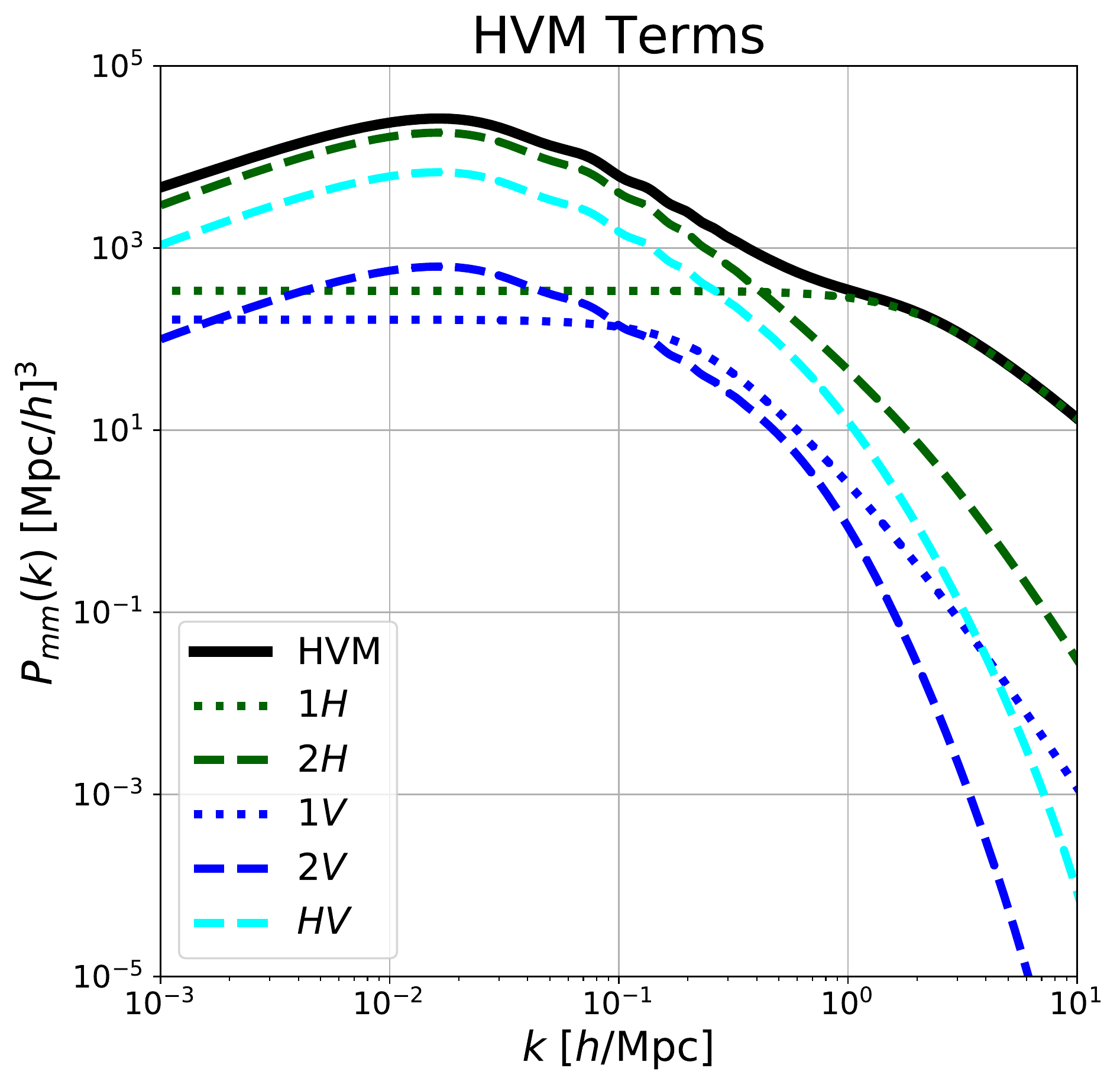}
\includegraphics[width=0.49\textwidth]{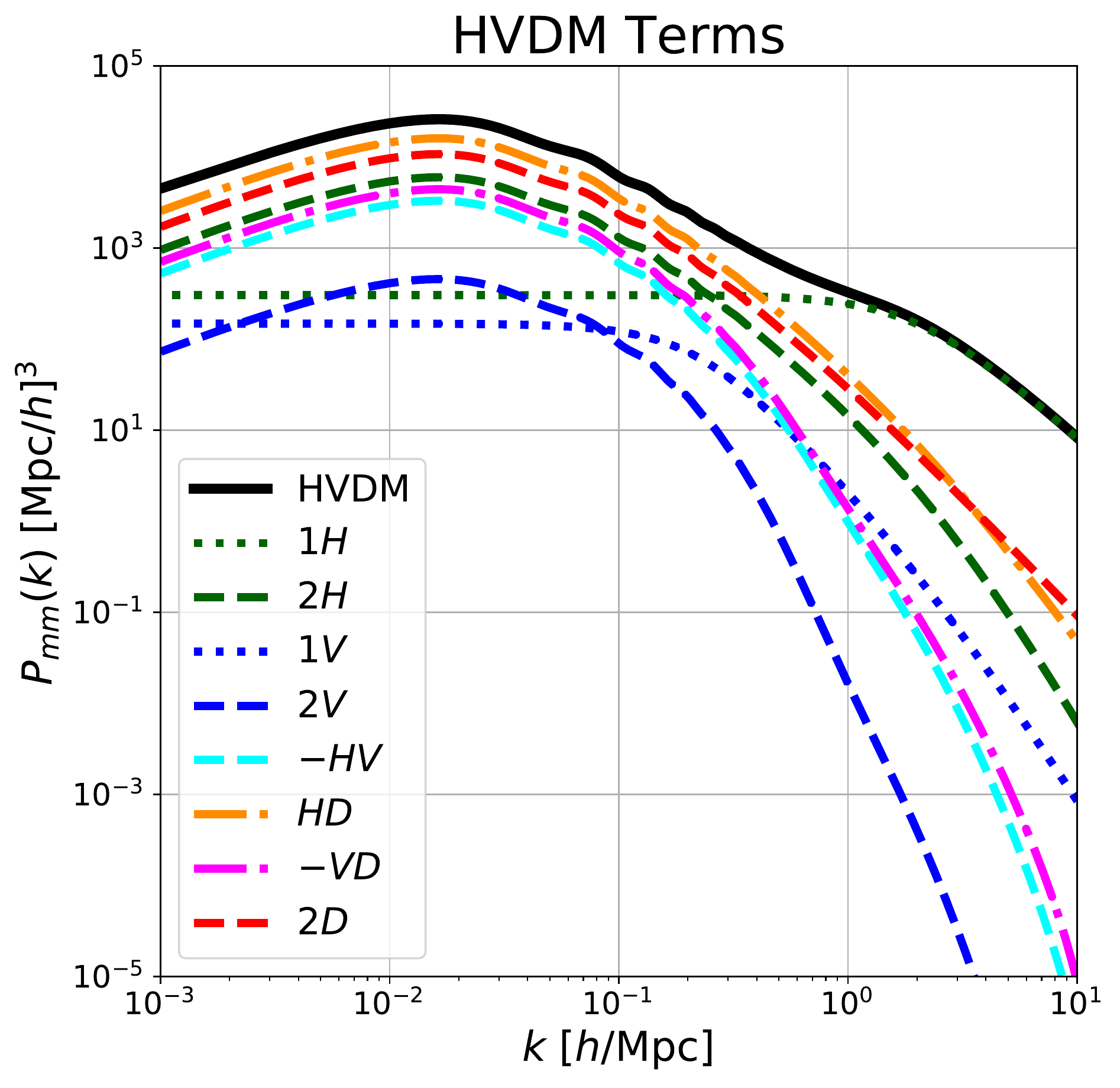}
\caption{\label{fig:hvm_terms}%
\small Halo Void Model All terms contributing to the matter-matter power spectrum $P_{mm}(k)$ in the Halo Void Model ({\it left}) and in the Halo Void Dust Model ({\it right}). The sum of all terms is shown as the solid black lines. Terms that contribute mainly on large scales are shown in dashed lines and terms that contribute mainly on small scales are shown in dotted lines. }
\end{figure}

For the matter-matter power spectrum, the HVM and the HVDM corrections are described respectively by Eqs.~(\ref{eq:Halo_Void_Model}) and (\ref{eq:Halo_Void_Dust_Model}).
We show each of the HVM (HVDM) terms in the left (right) panel of Fig.~\ref{fig:hvm_terms}. 

For the HVM, the original terms present in the HM are dominant at all scales and the void contributions are sub-leading, as expected since the mean bias of matter in voids is only of $0.15$ (see Fig.~\ref{fig:abundances_min}). This makes the contribution coming from halos at large scales almost six times larger than that coming from voids. The main correction comes from the $P_{mm}^{HV}$ term, which gives corrections $\mathcal{O}(10\%)$.
Notice that $P_{mm}^{HV}$, $P_{mm}^{2H}$ and $P_{mm}^{2V}$ have quite similar shapes at very large scales, and that is the reason why in the end both HM and HVM agree on those scales -- despite their physics being different, and the fact that the HM needs to be normalized. On smaller scales, the three terms start to differ, as seen in  the left panel of Fig.~\ref{fig:hvm_terms}, and their sum will be different. 

For the HVDM, the dust terms provide leading-order contributions, since it corresponds to $65\%$ of the total matter bias. $P_{mm}^{2H}$, $P_{mm}^{2V}$, $P_{mm}^{HV}$, $P_{mm}^{2D}$, $P_{mm}^{HD}$ and $P_{mm}^{VD}$ are similar at large scales but differ for $ k > 0.1 h/{\rm Mpc}$.

\begin{figure}[ht]
\centering
\includegraphics[width=0.49\textwidth]{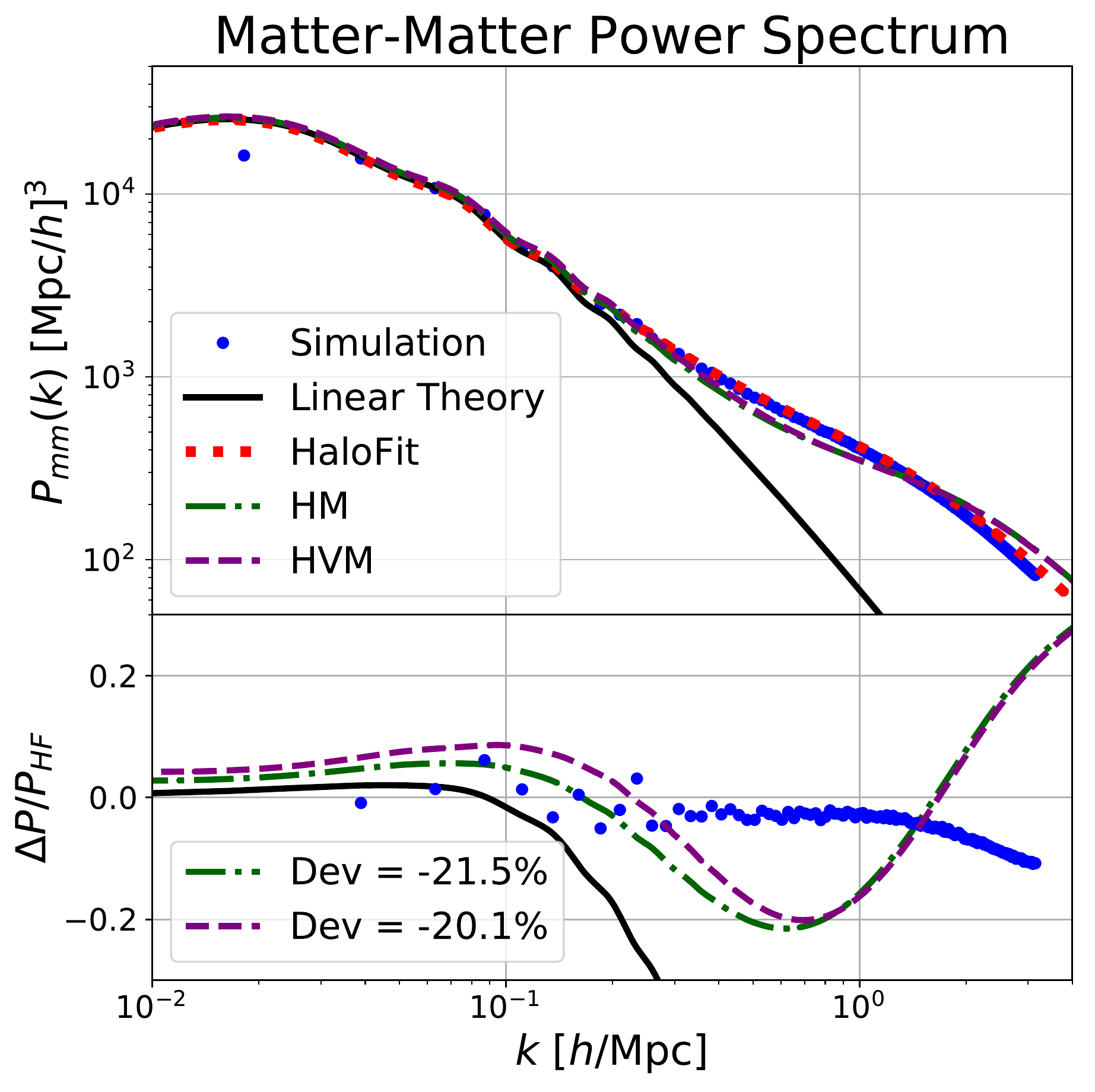}
\includegraphics[width=0.49\textwidth]{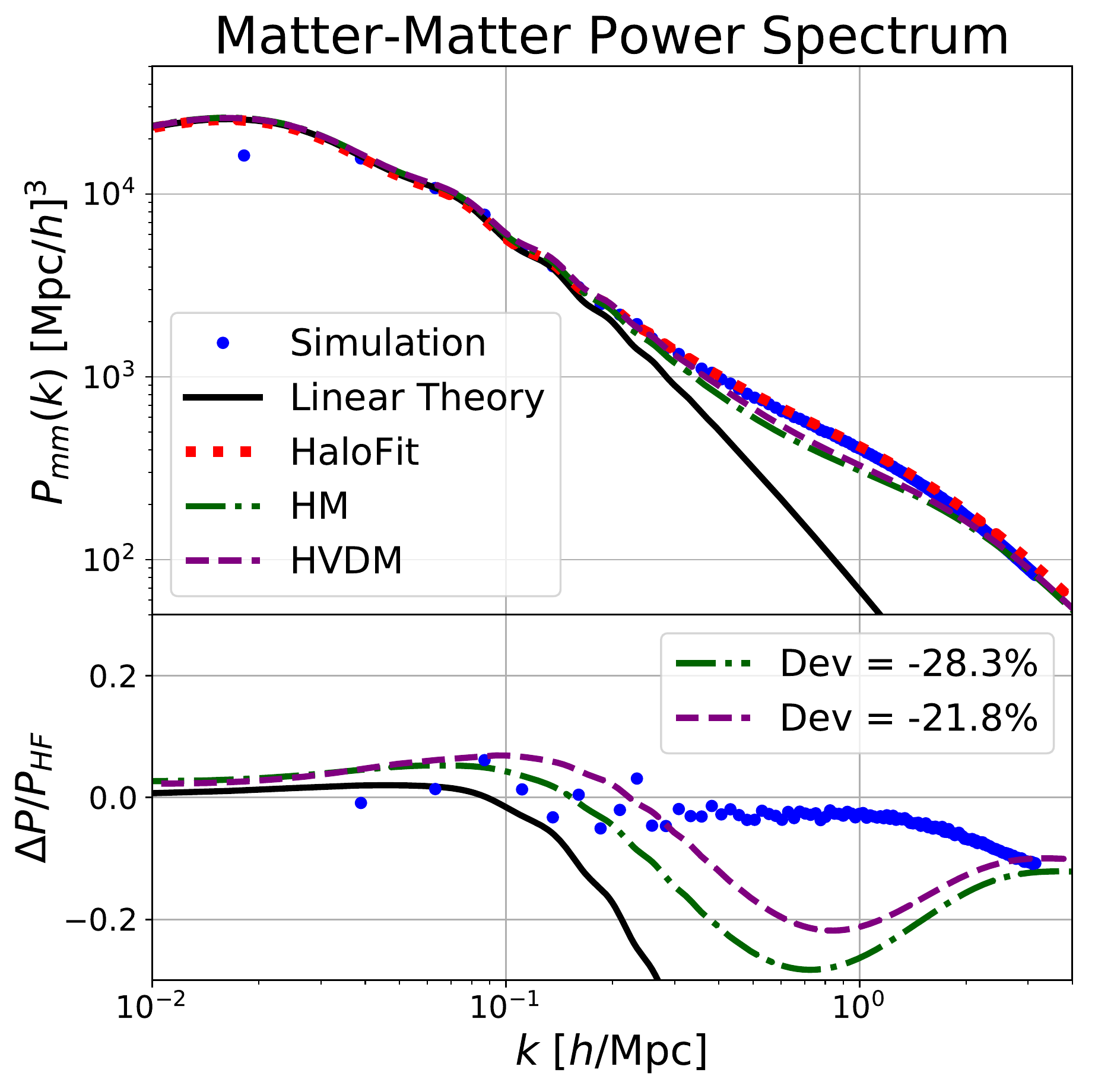}
\caption{\label{fig:mm_result}%
\small Matter-matter power spectrum using the HVM prescription ({\it left}) and the HVDM  prescription ({\it right}). The bottom panel shows the relative deviation of these models compared to the prediction from \texttt{HaloFit}. Notice that the HVM improves a few percent the HM prediction. Considering dust, the improvement is larger than $6\%$. At both the small and large $k$ limit, the HV(D)M agrees with the HM. Notice that the HM prediction changes sensibly as we consider different ingredients (left versus right).
}
\end{figure}

In Fig.~\ref{fig:mm_result} we also compare the different models with the fit to simulations from \texttt{HaloFit} \cite{Takahashi}.  We can see from the left panel that the transition from the 2Halo to the 1Halo term is slightly improved for the HVM compared to the HM (the maximum error in the transition of the 2Halo to the 1Halo term going from $21.5\%$ to  $20.1\%$). For the HVDM, the improvement is even better -- from $28.3\%$ to  $21.8\%$. In both cases the inclusion of the extra terms did not change the behaviour at large and small scales. At small scales, the HM error is very dependent on the prescriptions used and our results are consistent with other works in literature \cite{Mead}. Notice that even the HM prediction changes when we consider either the HVM or the HVDM prescription, because of the different bias and mass function parameters in both cases. Since the ingredients of the HVDM are less constrained, the HVDM works better at smaller scales\footnote{Looking only at the maximal deviation between the 2Halo and the 1Halo term, one could think that the HVM works better in that case, but this is not a good overall metric. This only happens because the 1Halo term is overestimated for the HVM, such that it helps correct the transition scales but jeopardizes smaller scales.}.

We must emphasize that the errors in the matter-matter spectrum could be severely reduced using other prescriptions (e.g. using Sheth \& Tormen mass function and linear bias for halos) or using different values for the halo and void overdensities (e.g. using $\Delta_{h} = 200$ for halos). When we choose a different combination of ingredients and values, in a manner that is not fully self-consistent, we can reach an accuracy of about $11\%$ for the HM and $5\%$ for the HVDM, showing that the $6\%$ improvement of the HVDM remains. However, the main goal of this work is not to obtain the best overall result, but using a framework that allow us to compare the HM to the HV(D)M and that correctly fixes the normalization on large scales.

We also highlight that the slight overprediction in the matter power spectrum on large scales ($k\lesssim 0.2$) in the HM and HV(D)M occurs because of the 1Halo (and 1Void) term(s) that become a shot-noise-like term for small $k$'s. This term can be suppressed using the formalism developed in \cite{Schmidt2} and therefore is not a relevant issue.

\subsection{Halo-matter spectrum and observed halo profile}
\label{sec:mh_results} 

As mentioned in Sec.~\ref{sec:halo_void_model}, one of the main features of the HM is its ability to make predictions also for cross-correlations. The HVM correction for the halo-matter power spectrum is given by Eq.~(\ref{eq:HVM_phm}) and the dust extra term by Eq.~(\ref{eq:HVDM_phm}).

\begin{figure}[ht]
\centering
\includegraphics[width=0.49\textwidth]{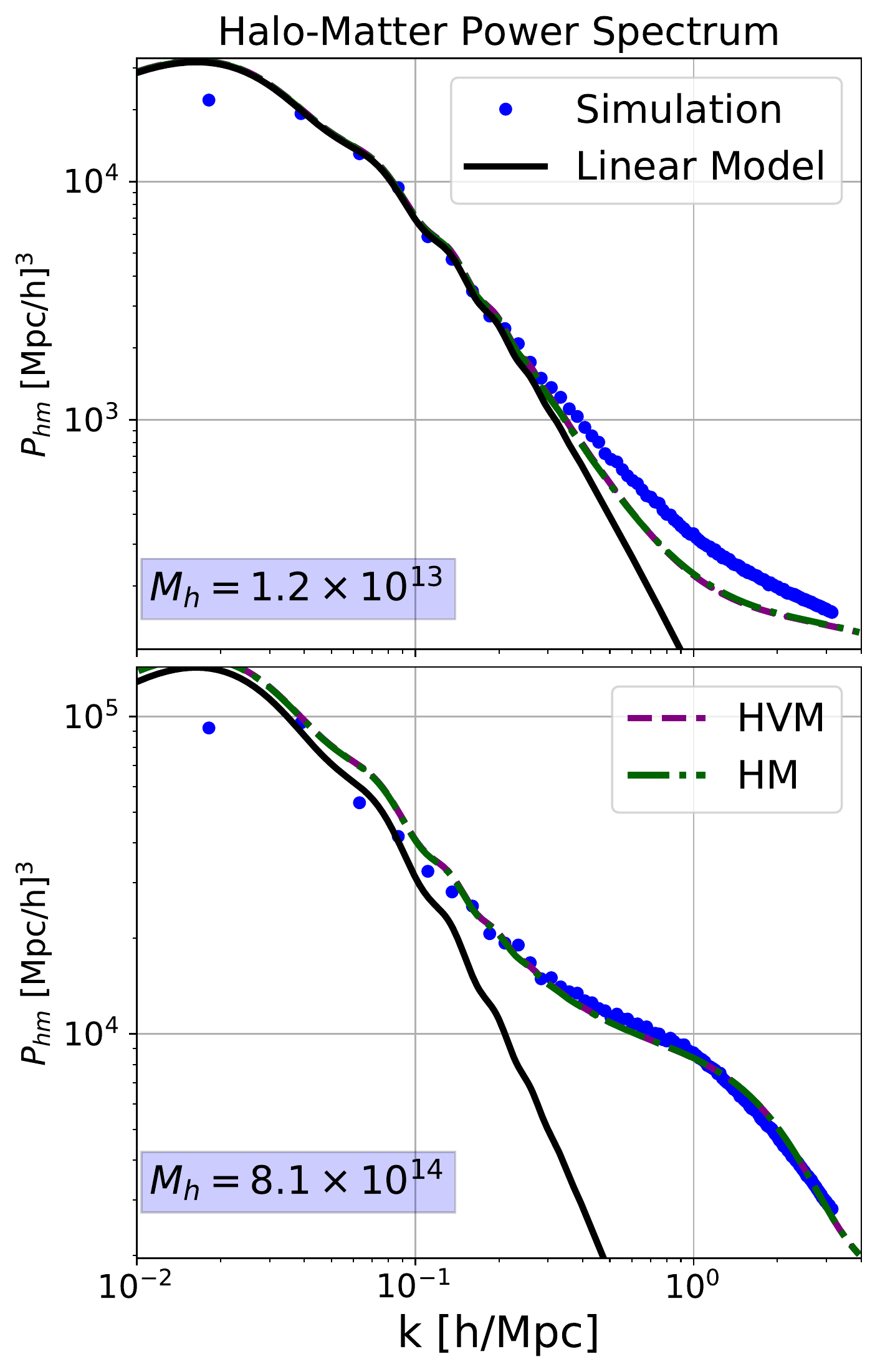}
\includegraphics[width=0.49\textwidth]{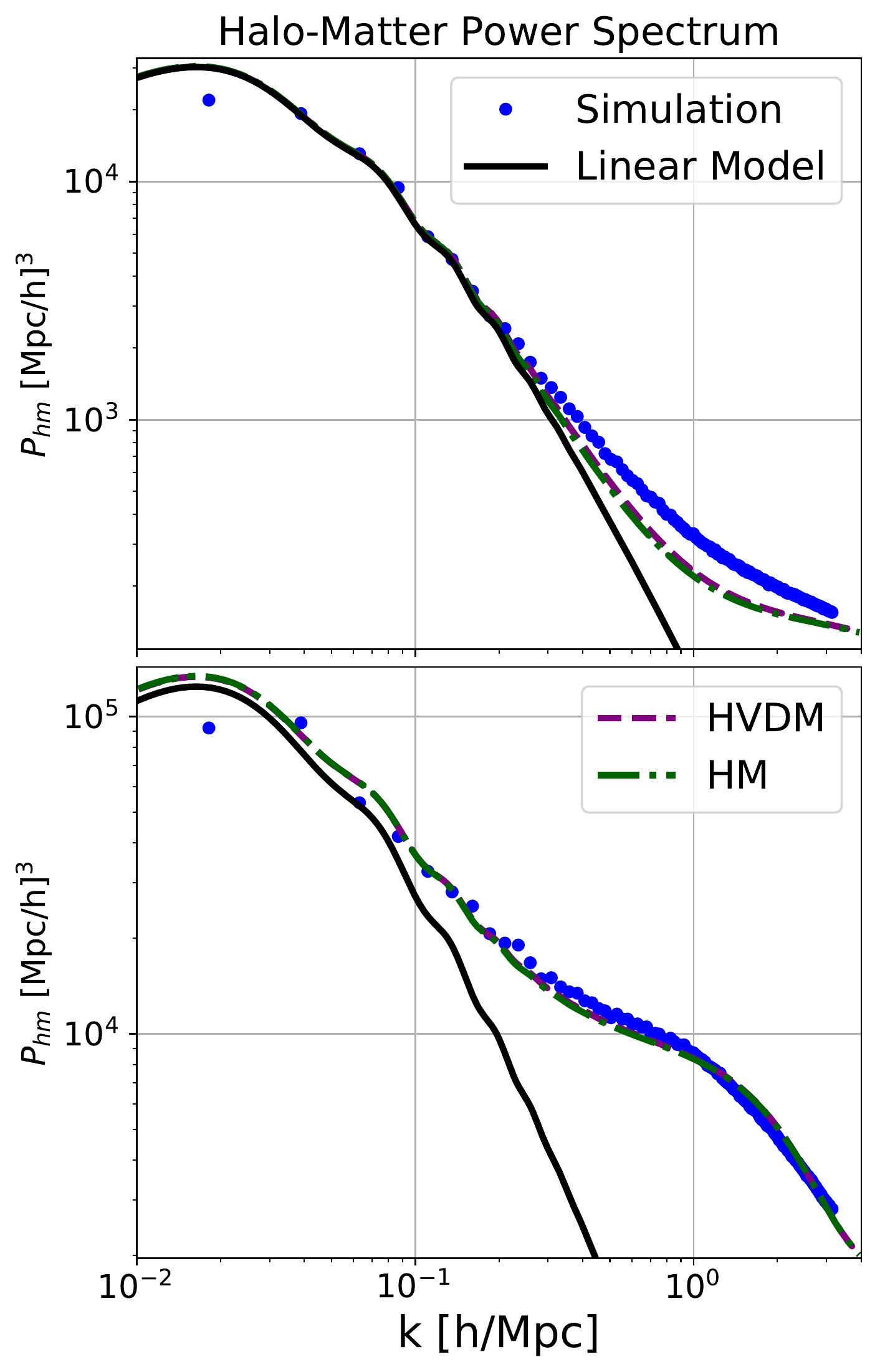}
\caption{\label{fig:Phm} \small Halo-matter power spectrum for two different bins of masses (in units of $M_{\odot}/h$). In the left, we compare the Halo Void Model (purple dashed) prescription with the Halo Model (green dashed-dots) and the linear model (red continuous). Notice that the HVM and the HM lines are indistinguishable and they differ from each other in less than $1\%$ here. We can see that both models predictions improve for large halo masses. In the right, the same but considering the Halo Void Dust Model.}
\end{figure}

In the left panels of Fig.~\ref{fig:Phm}, we display the difference between the HM and the HVM in the halo-matter power spectrum for two bins of mass.
Void and dust contributions to halo statistics are expected to be small compared to the self contribution of halos, which dominate the matter component. For instance $P_{hm}^{HV}$ only gives a small contribution, and the HVM reproduces the HM with marginal improvement. We can see that for small halo masses, the typical 1Halo term scale is smaller, such that this term does not correct the linear power spectrum by the right amount at large $k$. For large mass halos, both the HM and the HVM predictions reproduce better the simulated data. The results for the HVDM (right panels of Fig.~\ref{fig:Phm}) are similar.

\begin{figure}[ht]
\centering
\includegraphics[width=0.49\textwidth]{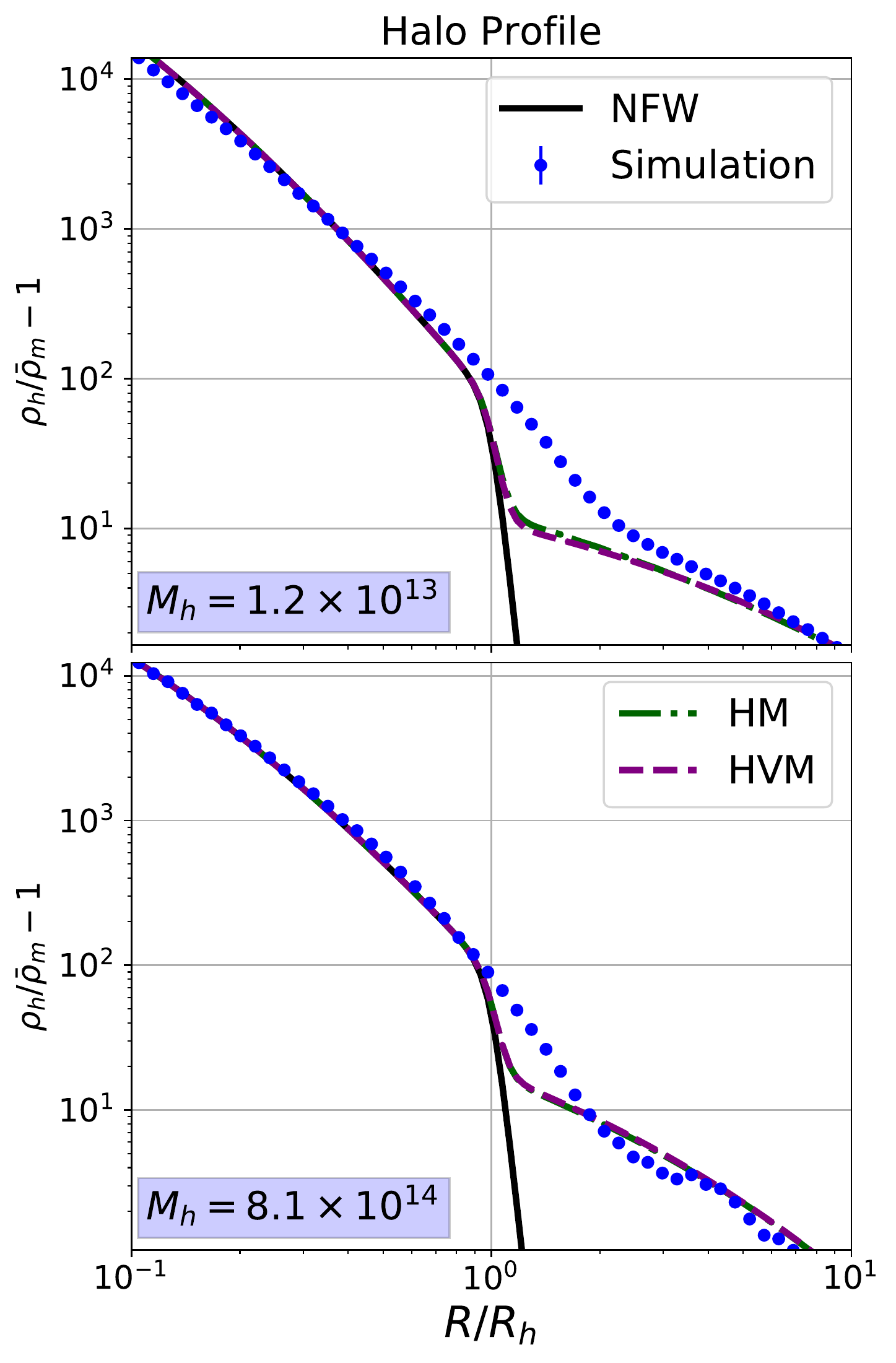}
\includegraphics[width=0.49\textwidth]{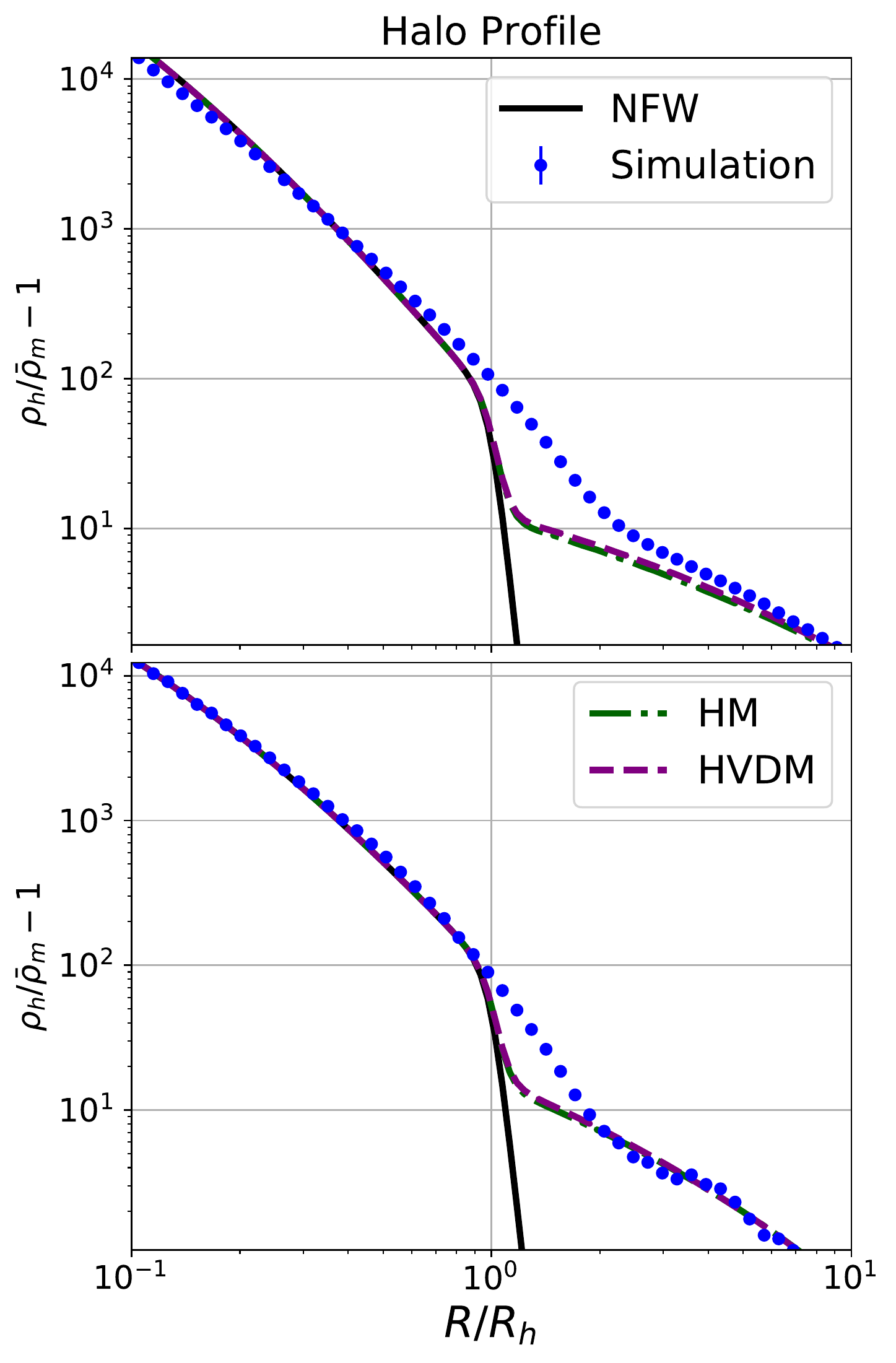}
\caption{\label{fig:Phm_hvdm} \small Halo-matter cross-correlation function or observed halo profile for two mass bins (in units of $M_{\odot}/h$). Results are shown for the HVM ({\it left}) and for the HVDM ({\it right}). The difference between these and the standard HM (green line) is marginal. The corrections of the HM to the NFW profile (black continuous) are also present in both HVM and HVDM. We have used the same ingredients both in the HM as in the HV(D)M, as described in Sec.~\ref{sec:results}.}
\end{figure}

The Fourier transform of $P_{hm}$ gives the halo-matter correlation $\xi_{hm}$ or the {\it observed} halo profile, which is shown in Fig.~\ref{fig:Phm_hvdm}. The 2Halo term of the Halo Model brings a relevant correction to the NFW profile (1Halo term) outside the halo radius (see left panel of Fig.~\ref{fig:profiles_hv}), where the 2Halo term fixes the profile to take into account contributions of the matter in other structures. Here we consider the truncated NFW profiles, as used in the HM computations, and show the large scales corrections on top of this truncated profile. We chose to show the profiles in this way to highlight the fact that we only need to have a correct prescription of the matter distribution around halos up to the halo radius.

The HVM adds corrections described by Eq.~(\ref{eq:xi_profile_h}) to the halo profile and the HVDM also gives an additional term Eq.~(\ref{eq:xi_profile_h_dust}).
Since the predicted spectrum of the HV(D)M differs from the HM halo-matter spectrum by $1\%$, the differences in the profiles are also negligible. As expected, we see that the HM and the HV(D)M work better for the large mass halos in the halo-matter power spectrum.

Another interesting point that is made clear in Fig.~\ref{fig:Phm_hvdm} is the incorrect prediction of the splashback radius \cite{More, Diemer} by the HM\footnote{The splashback radius is generally defined as the radius where the slope of the profile $d\rho/dr$ changes behaviour. Within the HM it should be related to the radius of transition between the 1Halo and large scale terms that contribute to the obsved halo profile.}. Within the HM prediction, this change of behavior happens close to $r_{\rm vir}$, whereas the actual transition from the simulation profile is closer to $2r_{\rm vir}$. This result also indicates that the virial radius does not seem to be the best choice for defining the halo boundary. Even though the halo radius is somewhat arbitrary, using the splashback radius or an overdensity lower than $\Delta_{\rm vir}$ could improve the transition between the 1Halo and large scale terms.

\subsection{Void-matter spectrum and observed void profile} \label{sec:mv_results}

\begin{figure}[ht]
\centering
\includegraphics[width=0.48\textwidth]{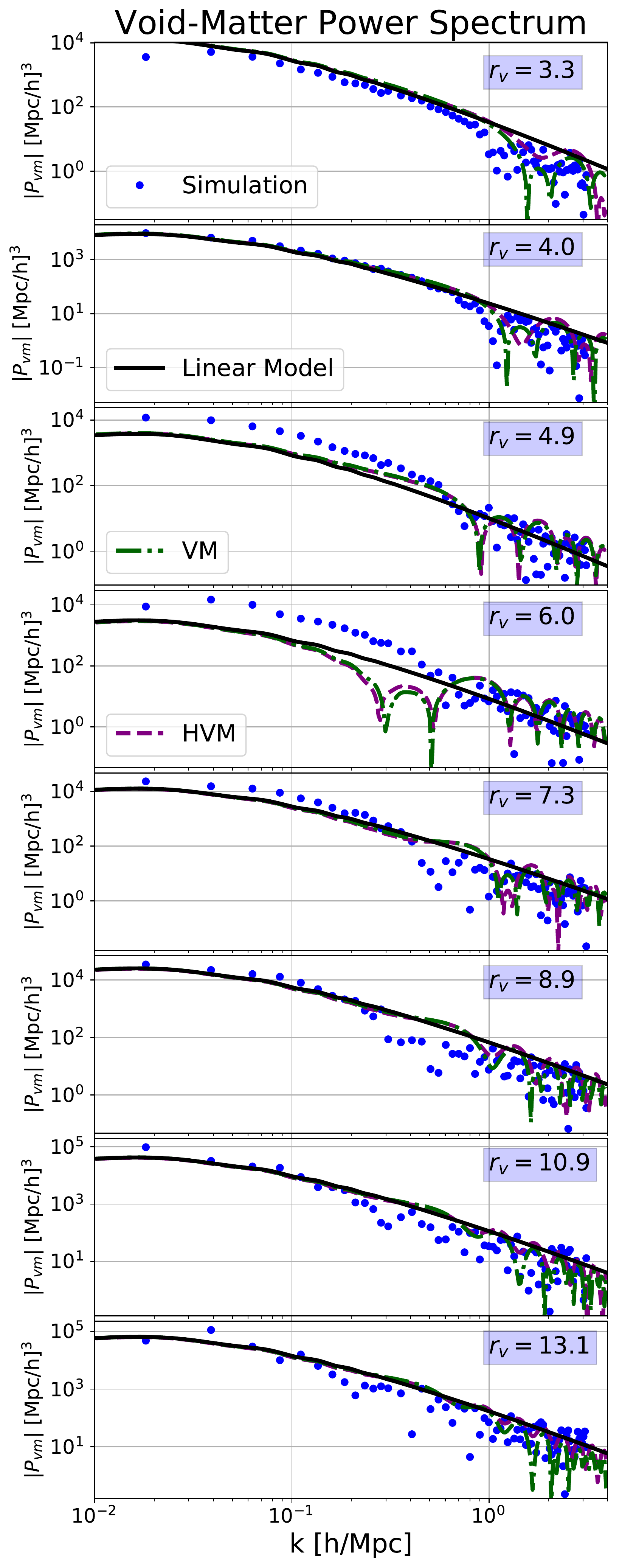}
\includegraphics[width=0.48\textwidth]{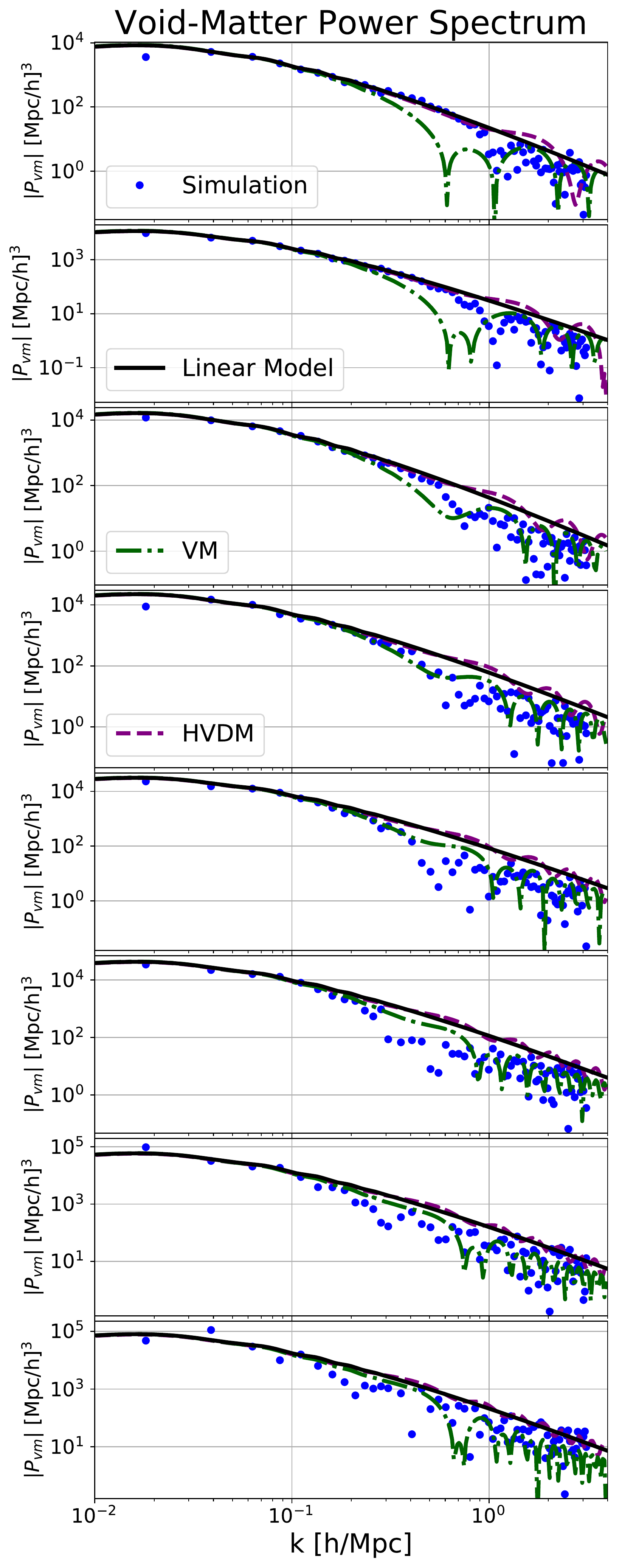}
\caption{\label{fig:Pvm} \small Void-matter power spectrum for eight different bins of void radius (in units of ${\rm Mpc}/h$). On the left panels, we compare the HVM and the so-called Void Model (see Eq.~(\ref{eq:VM_pvm})). The Void Model is a simplified version of the HVM that we propose for voids. Notice that since the linear bias is more constrained in that case by imposing the constraint $\bar{b}_{m}^{d} = 0$, the errors for some radius bins are larger. On the right panels, we compare the HVDM to the Void Model. Notice that now the void bias is better estimated and theory reproduces data at large scales for all radii bins. }
\end{figure}

The void-matter power spectrum is given by Eq.~(\ref{eq:HVM_pvm}) and the extra dust term by Eq.~(\ref{eq:HVDM_pvm}). We show the HVM predictions for the void-matter spectrum for the eight bins of void radius (see Table~\ref{table:Mass_bins}) in the left panels of Fig.~\ref{fig:Pvm}. The HVDM results are in the right panels of the same figure. 

For the case of voids, clearly the HM makes no predictions. On the other hand, if we consider large scales where $u_{h}(k|M) = u_{v}(k|R) \rightarrow 1$, we can simplify the 2Void, Halo-Void and Void-Dust terms and write down a Void Model prediction
\begin{equation}\label{eq:VM_pvm}
    P_{vm}^{\rm Void\, Model}(k|R) = P_{vm}^{1V}(k|R) + b_{v}(R) P_{mm}^{L}(k) \,,
\end{equation}
where $P_{vm}^{1V}(k|R)$ is the usual 1Void term present in the HVM and HVDM.
This expression is similar to the one presented in \cite{Chan} and is much simpler to compute, since we only need to know the internal void density profile and the void linear bias.

In Fig.~\ref{fig:Pvm} we show the void-matter power spectrum for the eight bins of Table \ref{table:Mass_bins}. We display the measurements, the linear theory, the Void Model as defined above and show results for the HVM (in the left panels) and the HVDM (in the right panels). The models in the right column describe better the simulation because the fit used for the linear bias of the voids is better. We can see that both the HV(D)M and the Void Model predict an inversion in the sign of the power spectrum for small scales, which is also present in the measurements. The power spectrum from the simulation is very noisy at small scales because of significant shot-noise.

\begin{figure}[ht]
\centering
\includegraphics[width=0.48\textwidth]{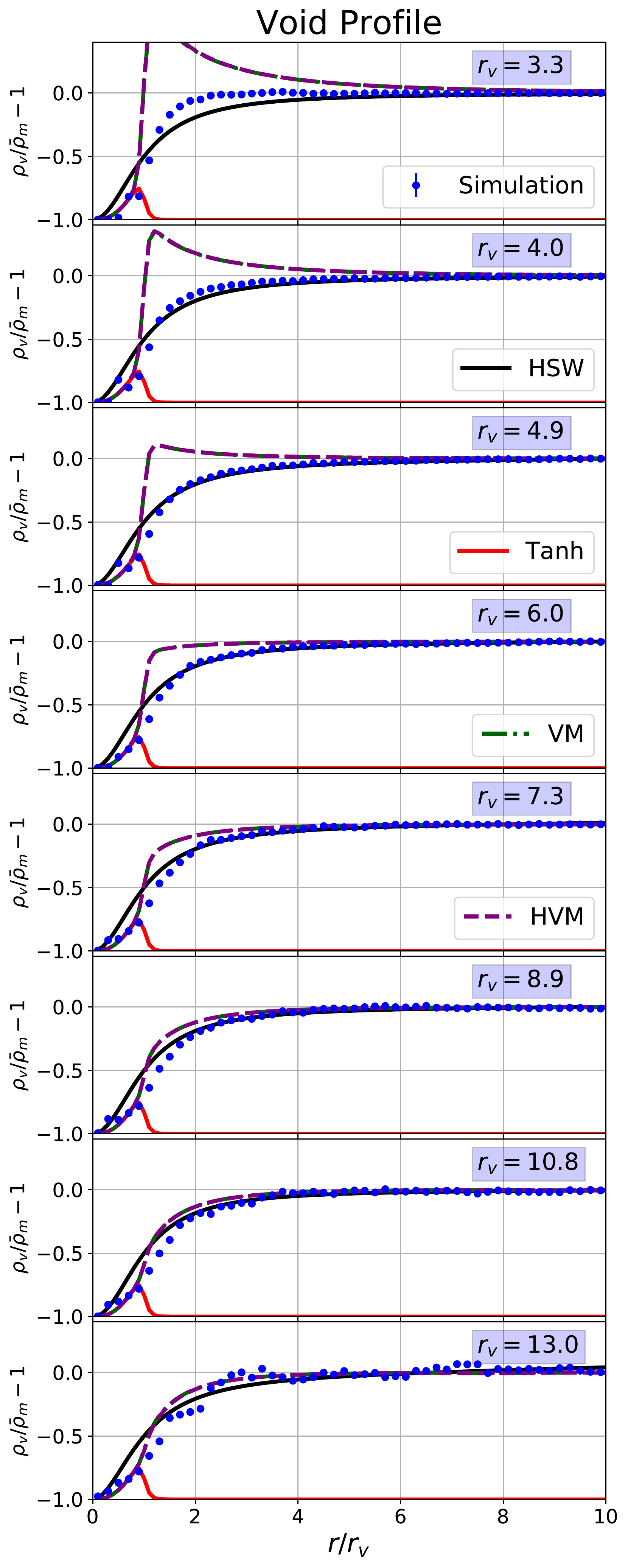}
\includegraphics[width=0.48\textwidth]{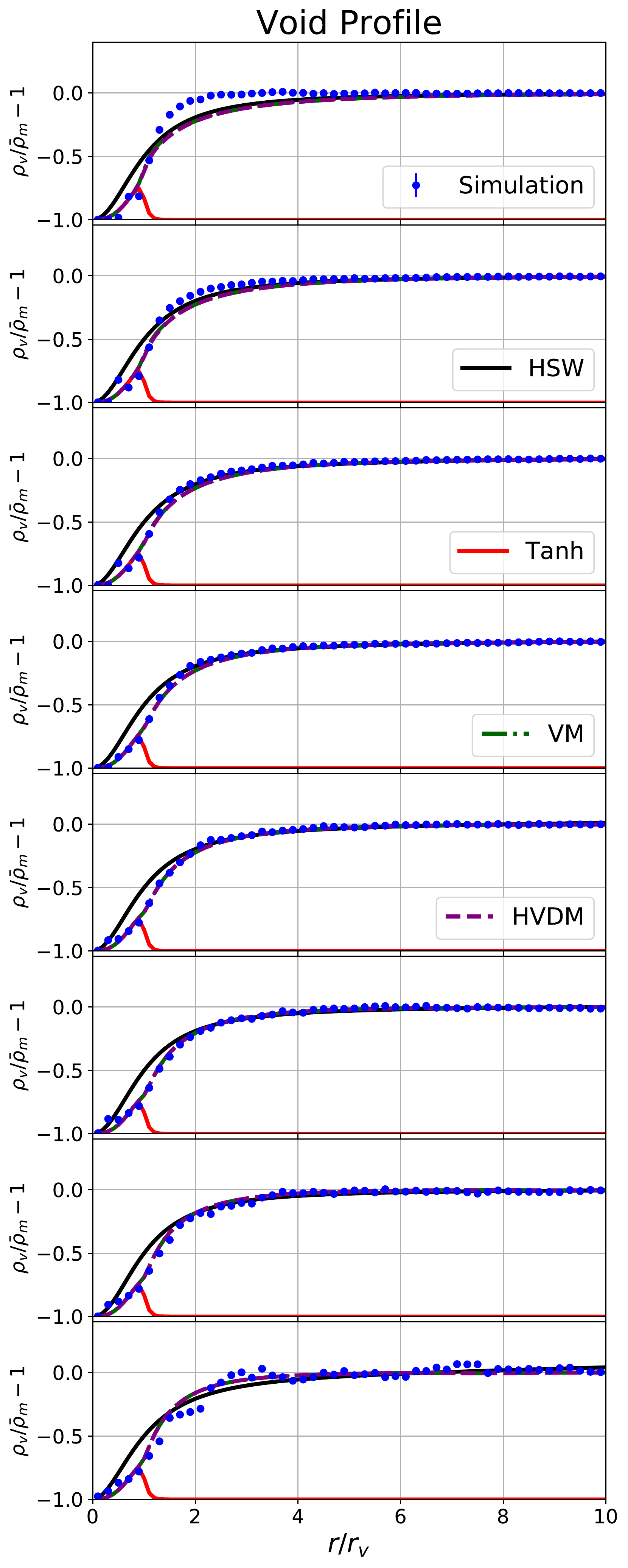}
\caption{\label{fig:HVM_Void_Profile} \small  HV(D)M prediction for the {\it observed} void profile for eight different void radius (in units of ${\rm Mpc}/h$). The HVM, on the left, corrects the $\tanh{}$ profile in Eq.~\eqref{eq:void_profile} to reproduce the background density for $r > r_v$. Notice that, as in Fig.~\ref{fig:Pvm}, the incorrect prediction of the void bias in the HVM for small void radii jeopardizes the profile prediction at those scales. On the right, the same is shown for the HVDM. Since the void bias has more freedom in the HVDM, it fixes the small radii bins keeping a good agreement for larger voids.}
\end{figure}

In Fig.~\ref{fig:HVM_Void_Profile} we show the {\it observed} void profile including the exclusion term described in Sec.~\ref{sec:bias}. The left panels refer to the HVM predictions. Notice that for small radius, in the left column, the fact that the void bias is fixed such that $\bar{b}_{m}^{d} = 0$ jeopardizes the theoretical prediction for all models.  For the HVDM, in the right panels, the small radius prediction is improved since the bias is corrected at these scales, as we can see in Fig.~\ref{fig:bias}. This improvement is the same shown in the power spectra. Both HVM and HVDM fix the $\tanh{}$ profile~(\ref{eq:void_profile}) in the region $r > r_v$, such that it reproduces the correct transient behavior to the background density. 

For the density profiles we do not find significant differences between the HV(D)M and the Void Model, showing that the simpler effective expression in Eq.~(\ref{eq:VM_pvm}), re-derived in this work, is enough to describe the matter distribution around cosmic voids.

\section{Conclusions}
\label{sec:conclusion}

In this work, we have modified the standard Halo Model of structure formation to include cosmic voids and dust matter in its formalism. Voids are relevant to probe multiple cosmological scenarios, especially to test gravity \cite{Voivodic1}.  
In order to compute predictions for the Halo Void (Dust) Model, we provide prescriptions for halo and void profile, abundance and linear bias in a self-consistent way. 

For the void profile, we develop a new fitting formula Eq.~(\ref{eq:void_profile}), with a single free parameter that better suits the void finder used in this work. 
For computing the abundance and bias, we consider two linear diffusive barriers (2LDB) in the excursion set formalism. We provide a new prescription for both halo and void bias, using 2LDB in the excursion set. This bias prescription seems to reproduce simulation data much more accurately than other models in the literature, especially for voids.

We show that considering halos and voids in the double barrier excursion set formalism immediately fulfills the constraint that all matter is either in halos or voids ($\bar{b}_{m}^{d} = 0$).  It dismisses any kind of artificial normalization or extrapolation for the 2Halo term, which is typically needed with the standard Halo Model. In addition, we show that the convergence of integrals containing the mass function and bias is much faster in this formalism (see Fig.~\ref{fig:abundances_min}), demonstrating that the Halo Void Model is a better effective description of how matter is distributed on large scales. Moreover the Halo Void Model is a self-consistent framework for the joint analysis of matter, halo and void correlation functions. When combined with e.g. Halo Occupation Distribution (HOD) and Void Occupation Distribution (VOD) models, we expect it to also describe statistics for galaxies, galaxy clusters  and galaxy voids. 

Finally we compare the 2-point statistics of the Halo Void (Dust) Model to the usual Halo Model. For the matter-matter power spectrum, the transition from the 2Halo to the 1Halo term improves by $1.4\%$ for the Halo Void Model and by $6.5\%$ for the Halo Void Dust Model. For the halo-matter spectrum and for the halo profile, the improvement is somewhat marginal. For the void-matter power spectrum and density profiles there is a large improvement with respect to the linear theory and we show that the large scale limit (i.e. the Void Model) is enough to predict these observables.

Similarly to the original Halo Model, within the HV(D)M framework it is also possible to compute higher order N-point statistics, covariance matrices and cross-correlations with the void density field. One could also construct a Halo Void Model for beyond $\Lambda$CDM models, in which voids could give larger contributions and play an important role. We leave these computations for future works.

Cosmology with voids is a fast expanding area with interesting perspectives to be explored, for which we expect this work to contribute. Our main point is that halos and voids are complementary structures of a coarse-grained description of the Universe. One of our key results is that, within the HV(D)M, it is possible to describe the innermost parts of the void density profile using just one free parameter (independent of the void radius). This should help improve cosmological analyses made with the void-galaxy correlation function and shed light on our understanding of gravity.

\acknowledgments
We thank the MPA infrastructure, in particular the FREYA cluster where the N-body simulation was run and analyzed.
RV is supported by FAPESP. 
HR is supported by the Deutsche Forschungsgemeinschaft under Germany's Excellence Strategy - EXC 2121 "Quantum Universe" - 390833306.
ML is partially supported by FAPESP and CNPq.

\appendix

\section{Excursion set predictions for halo/void abundance and linear bias}\label{app:abundance_bias}
    
    In this appendix we briefly review the excursion set theory \cite{Bond} computations of the mass function and linear bias. We closely follow \cite{deSimone} and \cite{deSimone2} to compute the conditional probability density distribution, using the formalism developed in \cite{Maggiore1, Maggiore2, Maggiore3}. We also use the same formalism to compute the unconditional probability density distribution. Given these two distributions we can compute the abundance and linear bias for both halos and voids. 
    
    As pointed out in the main text, we consider two linear diffusive barriers in order to account for cloud-in-void and void-in-cloud effects
    \begin{eqnarray}
    \left\langle B_{h}(S) \right\rangle = \delta _c + \beta _{h} S\,, \quad &\mbox{and}& \quad \left\langle B_{h}(S) B_{h}(S') \right\rangle = D_{h}\, \min(S, S') \,, \\
    \left\langle B_{v}(S) \right\rangle = \delta _v + \beta _{v} S\,, \quad &\mbox{and}& \quad \left\langle B_{v}(S) B_{v}(S') \right\rangle = D_{v}\, \min(S, S') \,,
    \label{eq:Barriers}
    \end{eqnarray}
    where $S$ is the variance of the linear matter density field defined by Eq.~(\ref{eq:variance}), $B_{h}(S)$ and $B_{v}(S)$ are respectively the barrier for the halo and void formation, $\delta _{c}$ and $\delta _{v}$ are the critical values for the formation of halo and void in the spherical evolution, $\beta _{h}$ and $\beta _{v}$ are the slopes of the linear dependence of the barrier with $S$, and $D_{h}$ and $D_{v}$ are the diffusion coefficients of each barrier.
    
    In Eq.~\eqref{eq:Barriers}, the parameter $\beta$ models the fact that the barrier for the formation of each structure should increase when the mass of that structure decreases, because smaller halos (voids) are less spherical and less likely to form. This happens since the three directions (axes) do not collapse simultaneously. The parameter $D$ models any residual uncertainty in the halo (void) definition, by changing the value of the spherical evolution prediction.
    Previous works have shown that the inclusion of these two parameters is enough to describe within $10\%$ precision the abundance of halos \cite{Corasaniti, Corasanitti2} and voids \cite{Voivodic1} in N-body simulations.
    
    Given the thresholds for halo and void formation, we compute the probability that the linear density contrast field smoothed at some scale has a value above (for halos) or below (for voids) the barriers.
    The density contrast field smoothed at some scale given by $S$\footnote{Note that in hierarchical models of structure formation there is a one-to-one relation between variance $S$ and mass $M$.} will be a solution of the Langevin equation
\begin{equation}
\frac{d \delta}{d S} = \eta (S) \,,
\label{eq:Langevin}
\end{equation}
where $\eta (S)$ is a white noise with zero mean and unit variance, if we consider a sharp-k window function to smooth the density field \cite{Maggiore1}. Note that the sharp-k window is the zeroth order contribution in the non-Markovian expansion of the sharp-x window.

    As shown originally in \cite{Maggiore1}, the probability density distribution ($\Pi$) to find an overdensity $\delta$ at some scale $S$ can be computed using the path integral formalism of the excursion set theory 
\begin{equation} 
\Pi(Y _{n}, S_{n}) = \int _{0} ^{\delta _{T}} dY _{1} \cdots \int _{0} ^{\delta _{T}} dY _{n-1} \int \mathcal{D} \lambda \ e^{Z} \,,
\label{eq:Path_Pi}
\end{equation}
where we defined $\delta _T = \delta _{c} + |\delta _{v}|$. The last integral above represents the probability density of one specific trajectory in Fourier space
\begin{equation}
\int \mathcal{D}  \lambda \ e^{Z} = \int _{-\infty} ^{+\infty} \frac{d\lambda _{1}}{2\pi} \cdots \int _{-\infty} ^{+\infty} \frac{d\lambda _{n}}{2\pi} e^{Z} \,.
\end{equation} 
For Gaussian perturbations
\begin{equation}
Z = i \sum _{j=1}^{\infty} \lambda _{j}Y _{j} - \frac{1}{2}\sum _{j, k = 1} ^{\infty} \lambda _{j} \lambda _{k} \left\langle Y(S_{j}) Y (S_{k}) \right\rangle \,,
\end{equation}
where we defined the new field $Y(S) = \delta (S) - B_{v}(S)$.
 
In order to write the upper limits of the integrals in Eq.~\eqref{eq:Path_Pi} in a simpler way, we also assumed $\beta _{h} = \beta _{v}$. This simplification has been shown to work well for voids \cite{Voivodic1}, but in principle one would have the same four free parameters for the halo and void excursions. We replace these four degrees of freedom by two barrier values for halo excursions and two for voids. Thi is justified by the fact that the top and bottom barriers decouple in the mass range considered here. 
	
	As shown in \cite{Corasanitti2}, this probability density $\Pi$ described by Eq.~(\ref{eq:Path_Pi}) is the solution of a modified Fokker-Planck equation
\begin{equation}
    \frac{\partial \Pi}{\partial S} = \frac{1 + D}{2}\frac{\partial ^{2}\Pi}{\partial Y^{2}} - \beta \frac{\partial \Pi}{\partial Y} \,,
    \label{eq:FP_eq}
\end{equation}
where $D = D_{h} + D_{v}$ and $\beta = \beta _{h} = \beta _{v}$.
    In order to consider the barriers and take into account the cloud-in-void and void-in-cloud effects we have to consider the boundary conditions
\begin{equation}
    \Pi(Y = 0, S) = 0 \,, \quad \mbox{and} \quad \Pi(Y = \delta _{T}, S) = 0\,,
    \label{eq:Boundary}
\end{equation}
and solve Eq.~(\ref{eq:FP_eq})  with the initial condition $\Pi(Y, S=0) = \delta _{D} (Y - |\delta _{v}|)$, which says that the density contrast is zero when we integrate over all the Universe.

Solving Eq.~(\ref{eq:FP_eq}) with the boundary conditions in Eq.~\eqref{eq:Boundary} we obtain the following multiplicity function \cite{Voivodic1}
\begin{eqnarray}
f^{\rm 2LDB}_{x}(S) &=& 2(1+D_x) \exp \left[-\frac{\beta_x^2S}{2(1+D_x)}-\frac{\beta _x\delta _x}{1+D_x}\right]  \nonumber \\
&\times & \sum_n \frac{n\pi}{\delta _T^2}S\sin \left( \frac{n\pi\delta _x}{\delta _T} \right)\exp \left[-\frac{n^2\pi^2(1+D_x)}{2\delta _T^2}S\right] \,,
\label{eq:f2LDB}
\end{eqnarray}
where the label $x = h$ for halos and $x=v$ for voids.
    Expression~\eqref{eq:f2LDB} corresponds to the 2LDB model used to predict the halo and void abundance in the main text (see Sec.~\ref{sec:mass_func}).
    
    Besides the unconditional density probability computed solving Eq.~\eqref{eq:FP_eq} with the initial condition $\Pi(Y, S=0) = \delta _{D} (Y - |\delta _{v}|)$, we can also solve this equation with the initial condition  $\Pi(Y, S=S_{0}) = \delta _{D} (Y - |\delta _{v}| + \delta _{m})$. We can then compute the density probability to form a halo or a void, assuming some overdensity $\delta _{m}$ at some larger scale $S_{0}$. Then, with this probability density we can compute the Lagrangian bias through the relation \cite{deSimone}
    \begin{equation}
    1 + \delta _{x}^{L} = \frac{f_{x}(S|\delta _{m}, S_{0})}{f_{x}(S)} \,,
    \end{equation}
    where we need to Taylor expand in $\delta _{m}$ and consider large scales where $S_{0} \gg S$. Here $f_{x}(S) = f_{x}(S|0,0)$ is the unconditional first crossing rate.
    
    By translational invariance of the Markovian random walks, we can easily write the expression for the conditional first crossing rate
    \begin{eqnarray}
    f^{\rm 2LDB}_{x}(S|\delta _{m}, S_{0}) &=& 2(1+D_x) \exp{\left[-\frac{\beta_x^2\Delta S}{2(1+D_x)}-\frac{\beta_x(\delta_x + \delta _{m})}{1+D_x}\right]}\nonumber \\ 
    &\times &\sum_n \frac{n\pi}{\delta_T^2}S\sin \left( \frac{n\pi(\delta_x + \delta _{m})}{\delta_T} \right)\exp{\left[-\frac{n^2\pi^2(1+D_x)}{2\delta_T^2}\Delta S\right]}\,,
    \label{eq:Conditional_2LDB}
    \end{eqnarray}
    where $\Delta S = S - S_{0}$.
    
    Expanding the expression in Eq.~\eqref{eq:Conditional_2LDB} up to first order in $\delta _{m}$, taking $S_{0} = 0$ (large scales), and normalizing by the unconditional first crossing rate in Eq.~\eqref{eq:f2LDB} we find
    \begin{equation}
    b_x^{\rm 2LDB}(S) = 1 \mp \frac{\sum_n \frac{n\pi}{\delta_T^2}\sin \left( \frac{n\pi \delta_x}{\delta_T}\right)\exp \left[-\frac{n^2\pi^2(1+D_x)}{2\delta_T^2}S\right]\left[ \mbox{cotan}  \left( \frac{n\pi \delta_{x}}{\delta _{T}}\right)\frac{n\pi}{\delta _{T}} - \frac{\beta_x}{1 + D_x} \right]} {\sum_n \frac{n\pi}{\delta_T^2}\sin \left( \frac{n\pi \delta_x}{\delta_T}\right)\exp \left[-\frac{n^2\pi^2(1+D_x)}{2\delta_T^2}S\right] } \,,
    \label{eq:b2LDB}
    \end{equation}
    where the plus sign is for halos and the minus sign is for voids.
    Eq.~\eqref{eq:b2LDB} is the 2LDB model used in the main text to model the linear bias of halos and voids in the computation of the HM, HVM and HVDM predictions.
    
\begin{figure}[ht]
\centering
\includegraphics[width=0.7\textwidth]{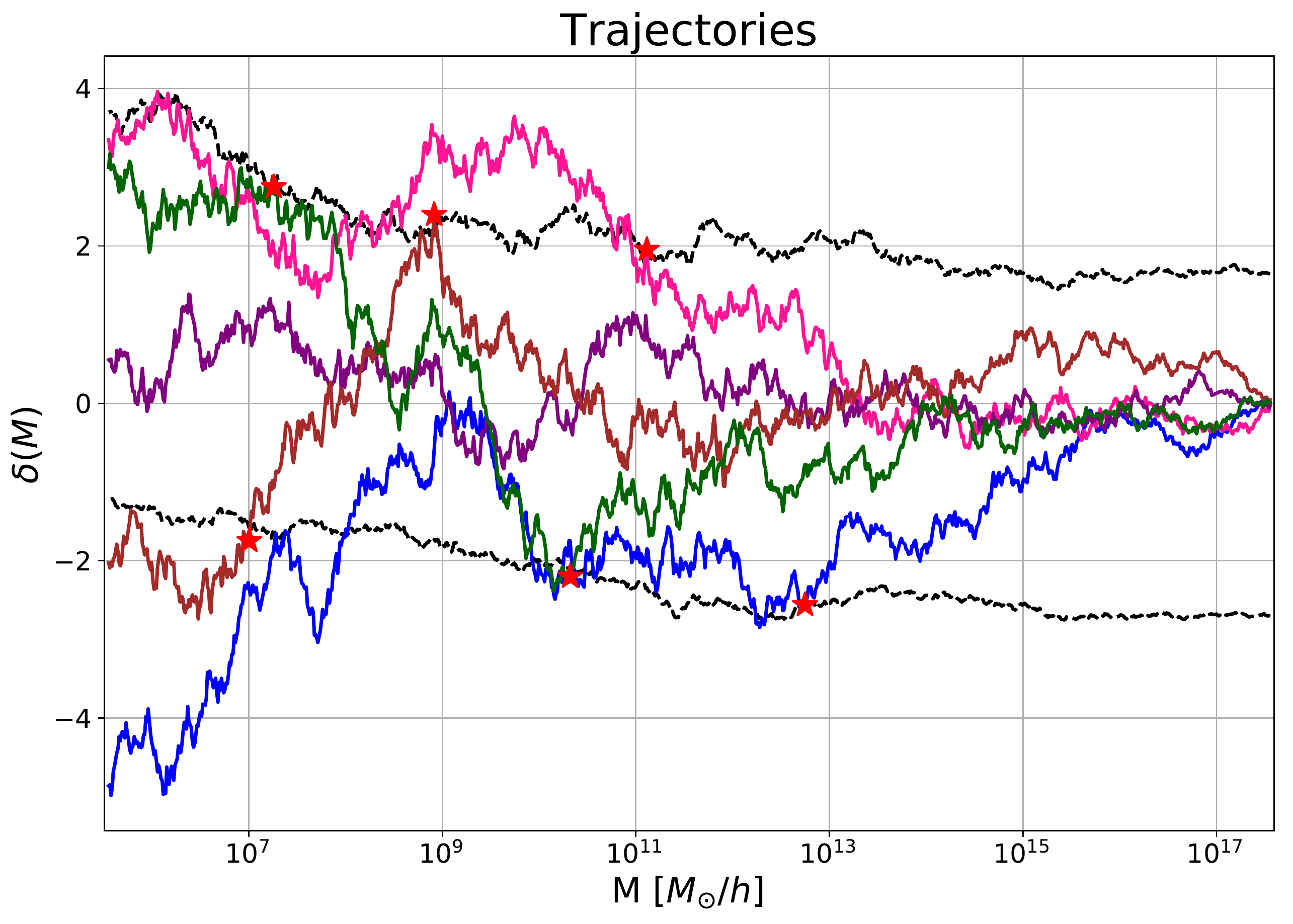}
\caption{\label{fig:Trajectories} \small Set with five different realizations of the trajectories for the density field and one realization of the two barriers using the best fits of the abundances. The black curves represent the two barriers and colored lines represent different trajectories. The red stars show the moment in which the trajectories cross the barriers for the first time.  The purple curve is the trajectory that have not crossed any barrier in this mass interval; the pink is the trajectory that have crossed just the halo barrier; the blue have crossed just the void barrier; the brown have crossed fist the halo barrier and then the void barrier and the green have crossed in the opposite order. Notice that the initial conditions are for $S = 0 $, which correspond to $M \rightarrow \infty$.}
\end{figure}

In Fig.~\ref{fig:Trajectories} we show an example of five different realizations of the density trajectory and the trajectories for the halo and void barriers. We see that the density trajectories cross the barrier more than once and, sometimes, cross both barriers (brown and green lines). This showcases the cloud-in-cloud (pink line), void-in-void (blue line), cloud-in-void (green line) and void-in-cloud (brown line) effects that need to be taken into account in order to avoid double counting of matter in the HVM. We can also see that some trajectories may not cross any barrier (purple line), indicating the importance of the dust term of the HVDM to properly account for all matter in the Universe.

\bibliographystyle{JHEP}
\bibliography{references}

\end{document}